Dennis Krämer[1], Anja Bosold[23], Martin Minarik[4], Cleo Schyvinck[5], André Hajek[6]

# Artificial Intelligence in Sports

## Insights from a Quantitative Survey among Sports Students in Germany about their Perceptions, Expectations, and Concerns regarding the Use of AI Tools

[1] University of Münster, Faculty of Education and Social Sciences
[2] Georg-August-University Göttingen, Faculty of Social Sciences
[3] Dennis Krämer and Anja Bosold contributed equally to the manuscript and are co-first authors.
[4] Georg-August-University Göttingen, Faculty of Social Sciences
[5] Ghent University, Department of Movement and Sports Sciences
[6] University Medical Center Hamburg-Eppendorf, Department of Health Economics and Health Services Research

**Contact**
PD Dr. Dennis Krämer
University of Münster
Faculty of Education and Social Sciences
Georgskommende 33, 48147 Münster, Germany
E-Mail: dennis.kraemer@uni-muenster.de
Web: https://www.uni-muenster.de/Soziologie/en/personen/kraemer.shtml

# Abstract

Generative Artificial Intelligence (AI) tools such as ChatGPT, Copilot, or Gemini have a crucial impact on academic research and teaching. However, empirical data on how students perceive the increasing influence of AI, which different types of tools they use, what they expect from them in their daily academic tasks, and their concerns regarding to the use of AI in their studies are still limited. The project “Artificial Intelligence in Sports” (AIS) aims to address this empirical gap through a quantitative study.

The manuscript presents findings from a quantitative survey conducted among sports students of all semesters in Germany using an online questionnaire. It explores aspects such as students’ usage behavior, motivational factors, and uncertainties regarding the potential impact of AI tools on academia in the future. Furthermore, the social climate in sports studies is being investigated to provide a general overview of the current situation of the students in Germany. Data collection took place between August and November 2023, addressing all sports departments at German universities, with a total of 262 students participating.

Our findings indicate that students have a strong interest in using AI tools in their studies, expecting them to improve their overall academic performance, understand the complexity of scientific approaches, and save time. They express confidence that the proliferation of AI will not compromise their critical thinking skills. Moreover, students are positive about integrating more AI-related topics into the curriculum and about lecturers adopting more AI-based teaching methods. However, our findings also show that students have concerns about plagiarism, lecturer preparedness and their own skills and future skill development.

In light of our empirical findings, we suggest that universities address this ambivalent situation by educating students on how to generate reliable information with AI and avoid misinformation. It is imperative that students have both practical and critical knowledge about the use and implications of a new technology that is rapidly spreading in academia. This will enable them to better understand the impact and also the limitations of a technology that only emerges through its practical use.

# 1
# Introduction

From a social science perspective, Artificial Intelligence (AI) represents the next step toward a digital society in which social interactions are increasingly influenced by technology-driven networks, automated processes, and accelerated communication. This process is driven by increasingly powerful devices, algorithms, and software, as well as by human actors and social institutions that support such developments. While current discussions about the impact of AI may seem recent, its roots can be traced back to the mid-20th century. At that time, AI was heralded as the "next big thing", capable of revolutionizing the analog world and surpassing human capabilities (Rammert, 2023). Its potential has been explored in various areas such as warfare, healthcare, and pandemics, but its impact on the education sector remains relatively underexplored (Ngo, 2023).

This has changed over the past two years. Media, public, and academic debates have increasingly addressed the potential of AI in areas such as self-driving cars, the transformation of social networking, and the reinforcement of social inequalities (Buolamwini & Gebru, 2018; Chmielewsky, 2018). Since the release of the generative AI tool ChatGPT in November 2022, this interest has extended to academic domains such as research, study, and teaching. Teachers, lecturers, and students have since faced the challenge of devising new criteria and assessment methods to revise established routines of information gathering, exchange, and performance evaluation. Within months of its release, ChatGPT had gained considerable traction, with millions of users worldwide integrating it into their daily work. In May 2023, OpenAI, the company behind ChatGPT, introduced the more powerful GPT-4, which can process not only text but also images in various forms, including photographs, diagrams, and screenshots.

The rapid pace of development makes it difficult to fully grasp the impact of AI. AI tools have the potential to support and transform the way knowledge is produced, structured, analyzed, synthesized, and ultimately understood. Current tools are able to translate and (re)formulate texts, generate outlines, identify relevant approaches and authors for specific research areas, generate ideas, mathematical formulas, programming codes, and compress complex information from books and papers within seconds. In empirical research, they can identify and create semantic structures in data. In theoretical research, AI tools can provide computational support for developing new approaches or generating interpretations.

However, these opportunities are accompanied by a number of risks, uncertainties, and social and ethical concerns. Many users are uncertain about the source of the generated knowledge, its reliability, its potential to create social inequalities and distortions, and whether the shared information will be further processed. In addition to concerns about the quality of the output generated, there are also questions about authorship, plagiarism, and misquotation.

Although the impact of AI tools on the education sector is currently a widely discussed topic, there is a lack of empirical research that examines the perceptions and practical uses of AI tools from the perspective of students. The extent to which students actually use AI tools, the opportunities and concerns they associate with AI, the support and skills they desire from lecturers, and the impact they anticipate AI will have on their academic, societal, and personal futures remain unclear. Against this backdrop, our study contributes to current research on AI in higher education by examining its role in academia from the perspective of students. It presents the results of a quantitative survey of sports students in Germany who completed an 18-page online questionnaire between August and November 2023.

Our interest in empirically exploring the perspectives of sports students is rooted in our academic affiliations. As researchers and lecturers in the social and sport sciences, we have observed over the past 18 months that many students are using AI tools for various tasks such as research, proofreading and translation. In our classes, we have often been confronted with the question of whether the use of AI tools, for example in writing a thesis, is permissible. On the one hand, students recognized the potential of AI tools, but on the other hand, they expressed uncertainty about whether and how to use them in their studies.

Interest in this area also stems from the fact that sport has been severely underrepresented in AI research to date. Recent studies show that generative AI tools are most frequently used by students of engineering (75.3%) and natural sciences (71.9%) (von Garrel, 2023). It is not yet empirically clear whether similar trends can be observed in the multidisciplinary and body-centered sports science.

In order to collect data, we contacted all universities in Germany that offer sports science degree programs. Many of them kindly supported us in distributing the online questionnaire to their students. In total, 262 students participated in our study. We would like to express our sincere gratitude for this invaluable support. Without their cooperation, this study and its findings would not have been possible. As our empirical research was privately funded, our goal is to expand our research to other countries and more study programs in the future, as our resources allow.

Dennis Krämer, Anja Bosold, Martin Minarik, Cleo Schyvinck, André Hajek

# 2
## Definition of terms: AI, GAI, NLP, LLM

> *"I am ChatGPT, an artificial intelligence language model developed by OpenAI. My primary function is to assist users like you by generating human-like text based on the input I receive. I have been trained on a diverse range of internet text up until January 2022, which allows me to understand and respond to a wide variety of topics and questions. While I strive to provide helpful and informative responses, it's important to note that I am a machine learning model and not a sentient being."*

This is a definition formulated by ChatGPT 3.5 in response to the prompt: "define yourself". It explains the fundamental principles of generative AI, a field that has transformed human-computer interaction and our understanding and use of AI. This chapter explores the key terms and definitions that form the basis of generative AI, examining its mechanisms and applications.

AI has been defined as "a system's ability to correctly interpret external data, to learn from such data, and to use those learnings to achieve specific goals and tasks through flexible adaptation" (Kaplan & Haenlein, 2019, p. 15). Generative AI (GAI) is a form of AI that not only interprets and learns from data but also produces original content in various forms, such as text, images, or videos (Dasborough, 2023). Generally, GAI can be classified into two types of mechanisms: unimodal models and multimodal models (Cao et al., 2023, p. 111:3). Unimodal models require instructions to be of the "same modality as the generated content modality, whereas multimodal models accept cross-modal instructions and produce results of different modalities" (ibid.). Unimodal models can only generate output in written language based on written instructions or prompts. This means that multimodal models have greater flexibility in terms of input and output modalities. In contrast, multimodal models allow users to provide written instructions while receiving results in the form of images or videos, and vice versa.

ChatGPT by Open AI, Google Gemini (formerly known as Bard), and Microsoft's Copilot are some of the most common examples of GAI that operate primarily through written or spoken language, using natural language processing (NLP). NLPs are forms of GAI that facilitate communication between applications and humans through natural language. Similar to human-human communication via messenger, NLP-based GAI simulates "conversations with human users through text-based interfaces." (ibid., p. 111:21). Users formulate prompts – queries, instructions, or requests – and receive human-like responses (Banh & Strobel, 2023; Ooi et al., 2023). This prompts-response exchange can be perpetuated in a conversational manner. Bozkurt and Sharma (2023) describe prompt engineering as a sophisticated craft that aims to "elicit specific types of responses or behaviors from an AI language model […] to optimize how the model responds based on the structure, content, and tone of the question, thereby facilitating more accurate, useful, or engaging interactions" (Bozkurt & Sharma, 2023, p. ii).

Large language models (LLM) are a specific type of architecture used to train NLP-based GAI. They utilize deep learning, a category of machine learning that employs large datasets from various sources, including the internet, to learn and mimic the formal criteria of natural language.

ChatGPT employs one such learning model, the generative pre-trained transformer (GPT) (Cascella et al., 2023). In applications such as ChatGPT, the automated learning process via the internet is endorsed by reinforcement learning from human feedback (RLHF). Human workers rate different answers given by the GAI to further feed and check the machine learning process (Cao et al., 2023, p. 111:6). This process of reinforced learning can also be carried out by other AI instead of human workers (Burtell & Toner, 2024b). In responses generated by large language models, the primary criterion is 'likelihood' rather than 'truth'. This means that when generating a sequence of words in response to human prompts, LLMs choose consecutive words based on their statistical likelihood of following the previous word (Ooi et al., 2023, p. 3). As a result, the sentences they construct may appear to be based on the accuracy of the information, but are actually based on the likelihood of word orders. This highlights some of the key issues with using GAI in information-based applications. While the information provided by ChatGPT and similar LLM-based GAI may appear reasonable, it is often unreliable in terms of accuracy (Cao et al., 2023, p. 111:8).

Additionally, there is a risk of circulating information that can be classified as "toxic" due to "biased outputs […], stereotypes […] and misinformation." (ibid., p. 111:27). However, it must be clarified that 'likelihood' cannot be considered the sole criterion, as the pure logic of 'next-word-prediction' only applies to the self-supervised pre-training stage of LLM-based GAI (Burtell & Toner, 2024a). Fine-tuning during the pre-training stage, as well as reinforcement through human workers and other AI models, provide another essential criterion that aims to eliminate the aforementioned flaws to which pre-trained GAI is prone.

However, Burtell and Toner (2024b) demonstrate that even with training and checking, unbiased and flawless solutions are not fully guaranteed. The main issues that persist are unexpected model behavior in new situations and the potential for "malicious actors" to corrupt the fine-tuning process. The authors mention that "'Jailbreaks', where models are 'tricked' into providing harmful or disallowed responses, is one example of this kind of circumvention. Another is research showing that models can be further fine-tuned by adversarial actors to undo their initial fine-tuning, or in some cases it can even be removed accidentally" (Burtell & Toner, 2024b).

# 3
## State of research

Research on AI has exploded recently due to current developments like the release of ChatGPT in November 2022. Von Garrel et al. (2023) refer to empirical work conducted in Germany in 2023 to depict the increase of AI adoption in the country. According to these scholars, utilization rates among the general population increased from 10% at the beginning of 2023, to 17% in March 2023, to 23% in May 2023. With regard to ChatGPT specifically, Neumann and colleagues (2023) report 200 million users worldwide, of whom 60% are aged between 18-34 years. Among these users, the majority are students and researchers (Neumann et al., 2023).

AI, including chatbots and other NLP tools, has been extensively studied in the field of education over the past 3 years (for reviews, see: Lo, 2023; Zawacki-Richter et al., 2019, Naz Ansari et al., 2023). In their scoping review, which included 69 studies, Naz Ansari et al. (2023) classified the majority of the papers as non-empirical, while only 27% of the studies were based on empirical data. According to the reviews by Zawacki-Richter et al. (2019) and Laupichler et al. (2022), the majority of the studies that empirically examine the role of AI in higher education employ a quantitative approach. Those quantitative studies (e.g., Bonsu et al., 2023; Chan, 2023; Raman et al., 2023; Sallam et al., 2024, Shoufan, 2023) mainly refer to ChatGPT as the most adopted AI tool. These studies utilized questionnaires to explore student perceptions and intentions to use ChatGPT (Bonsu et al., 2023), its benefits (Raman et al., 2023), as well as the major issues and risks concerning the use of AI in higher education (see Chan, 2023). For example, Raman et al. (2023) surveyed 288 university students about the benefits of adopting ChatGPT. They found that students view ChatGPT as innovative, compatible, and user-friendly, considering it as an innovative resource that enables the independent pursuit of educational goals. Consequently, the benefits provided by ChatGPT in education motivate students to utilize the tool. Alongside the positive features, risks and challenges of AI in higher education are receiving increased academic attention. Chan (2023) surveyed 457 university students across various disciplines in Hong Kong and found that students were concerned about over-reliance on AI technologies, limited social interaction, and the potential hindrance to the development of generic skills.

In a study amongst university students in the US, approximately one-third of the students surveyed had utilized ChatGPT to complete written homework assignments, with 60% using the program on more than half of their assignments (Intelligent.com, 2023). Similar findings were observed in Germany. A nationwide survey involving 6300 higher education students from over 10 faculties assessed usage behavior in the context of studies and everyday life. The results indicated that, on average, 63.2% of the participants use AI tools as part of their studies. For the sports faculty in particular, it was found that 87.5% of the students use AI-based tools (von Garrel et al., 2023). German students considered ChatGPT as the most important AI tool to support their learning. Similarly, research findings from the University of Düsseldorf, based on a sample of 304 students across faculties, showed the highest adoption of AI tools that support learning and provide recommendations, rather than tools that make decisions independently (Kieslich et al., 2019).

*Opportunities of AI in higher education*

Research has been conducted on the potential benefits of AI in enhancing student support in higher education (Lo, 2023). Lo et al (2023) identified two main applications, namely learning and assessment. In terms of learning, Dwivedi et al. (2023) and Geerling et al. (2023) highlighted benefits such as answering questions, summarizing books and articles, and writing essays. Concerning the assessment aspect, students benefit from using ChatGPT to prepare for exams, improve drafts of text, and seek assistance or feedback. For example, Gilson et al. (2023) noted that by asking scaffolding questions to AI, students applied their knowledge and reasoning skills, thereby improving writing and research skills. Research in Germany indicates that students' main motivation to use AI is to ask questions, seek clarifications, and improve understanding of subject-specific concepts (von Garrel et al., 2023). Sport-specific usage functions include translations (40.6%), text processing (37.5%), literature study (37.5%), and exam preparation (21.9%) (von Garrel et al., 2023).
Additionally, AI has received interest as a potential way to improve the educational experience for students. According to Kuhail et al. (2022), personalized instruction has been found to boost student performance, engagement, and confidence in their own abilities in the classroom. When students are provided with information tailored to their own interests and skill sets, they are more likely to actively participate and learn. Consequently, Khan et al. (2021) and Ghnemat et al. (2022) concluded that the utilization of AI tools can result in significant improvements in academic performance.

*Risks of AI in higher education*

While there is much to be gained from implementing AI in higher education, there are challenges and risks involved. A growing body of scientific works has begun to document the potential issues of AI in education (Kasneci et al., 2023; Lim et al., 2023). Lo (2023) identified four major issues associated with ChatGPT in education: Firstly, there are concerns regarding misinformation or fake information. Mogali (2024) raised concerns about ChatGPT's provision of inaccurate information, and Perkins (2023) observed that the bibliographic citations generated by ChatGPT can be fake. ChatGPT's accuracy and reliability have been found to be questionable in certain subject domains, such as mathematics (Frieder et al., 2023) and health science (Mogali, 2024), as well as sport science (Szabo, 2023). Secondly, there is a risk of relying on biased information. A study by Bolukbasi and colleagues (2016) demonstrated that generative AI models trained on large bodies of text from the internet can exhibit gender bias in their language generation. Generative models lack the ability to understand and reflect upon context and situations, which can result in inappropriate or irrelevant responses. Thirdly, there is a limitation in up-to-date knowledge. Many generative AI models can only provide information based on the data they were trained on up to a specific time, and lack access to more recent information (Rasul et al., 2023). The incorrect, biased, and incomplete information and references generated by ChatGPT could potentially mislead students (Hsu & Thompson, 2023). A fourth important issue is plagiarism. Evidence suggests that essays generated by ChatGPT can bypass conventional plagiarism detectors (Lo, 2023). Consequently, there have been attempts to ban AI tools like ChatGPT from educational institutions out of fear of negatively impacting education and learning. One third of the universities in the prestigious UK Russell Group, including Oxford and Cambridge, have declared the use of AI for assignments as academic misconduct.

To what extent students are aware of these risks and concerned about them remains to be further studied. According to Chan (2023), students do indicate some other risks. Firstly, there's a risk associated with the lack of human interaction. While ChatGPT can simulate conversations, it cannot replace the social and emotional interactions that occur in face-to-face interactions, which are particularly important for sports students, including body-to-body interactions. Secondly, over-reliance on AI tools can influence students' cognitive abilities and language skills (Choi et al., 2023). Thirdly, Sullivan et al. (2023) refer to concerns about privacy and data security when using ChatGPT and other generative AI models in education. Similarly, German scholars note that data protection and personal rights represent key concerns in the implementation of AI in higher education (Schmohl et al., 2023). Given that AI, with all its advantages and risks, is rapidly growing, additional research is needed to explore its sustainable integration in education.

# 4
# Methods section

Our interdisciplinary research team consists of five scholars with expertise in sport sociology, health research, and sport management. The main objective was to provide an empirical contribution to complement the current research landscape on the role and significance of AI tools in universities, with a particular focus on exploring the perspective of students. To gain an overview, the project group started with a comprehensive literature review of international empirical studies focusing on the role of AI tools in higher education. The central focus was on identifying relevant research findings, as well as existing questionnaires and scales used for data collection. Several initial team meetings were held to clarify how to effectively design the research project given the limited financial and time resources, and to achieve a high-level participation from sport departments and students in our study. In order to achieve this, we decided on a three-step approach: 1) to contact sports students through the sports departments of German universities, 2) to extend the study to universities in Belgium, and 3) to expand the study to other study degree programs and other European countries.

*Item development*

In order to develop the measuring instrument, existing items were first extracted and compiled from questionnaires exploring AI in academia (Bonsu & Baffour-Kodua, 2023; Chan & Hu, 2023; Chan & Zhou, 2023; Karaca et al., 2021; Raman et al., 2023; Sallam et al., 2023; Shoufan, 2023). From there, a total of 130 items were identified and thematically assigned to a preliminary set of constructs defined by the research group during several project meetings. This process resulted in 5 dimensions that addressed the question of the influence of AI: “Skills and Competencies”, “Motivation and Values”, “Vision and Opportunities”, “Risks and Concerns”, and “Inequality Considerations”. In developing the preliminary constructs, we referred to factors and subscales that were confirmed by factor analyses in the studies of the above-mentioned authors.

The next step was to screen the items within our research group and assess their relevance to the planned research. On this basis, 36 items were retained. In addition, items were consolidated for new empirical areas such as privacy and security, time-saving techniques, and professional skills. The items identified for inclusion in the questionnaire were transferred into a document, subsequently translated into German and adapted in their wording so that all items were consistent with the content related to the proposed target population (i.e., sports students) and the planned Likert scale. During the collaborative reflections within our research group, we decided to reformulate the initially defined constructs and to adapt them step by step during the research process. In this process, the constructs “Risks and Concerns” and “Inequality Considerations” were merged.
Furthermore, we discussed how AI should be defined in our study, considering the different definitions and meanings in academia. A precise definition was also important to avoid confusion among the surveyed students by informing them about the specific type of AI they should consider when answering the questionnaire. Given the current situation in academia, we decided to focus not on AI in general, but specifically on chat-based generative AI tools.
An important aspect that emerged during this process was that we needed to ask about students’ skills and experience in using AI tools at the beginning, since (non-)experience has a direct influence on follow-up questions. We addressed this with appropriate filtering solutions in the questionnaire.
One challenge was also to reformulate the items in a way that allowed both general questions about the importance of AI (e.g., the increasing influence of AI in studies) and specific questions about certain AI tools (e.g., the use of generative AI). To avoid methodological errors in the subsequent data analysis, we also introduced reverse-coded items.
During this process, we also worked on refining our previous sub-dimensions. We adapted the existing questionnaire by including additional specific questions, such as those about students’ usage patterns, evaluations of AI-generated results, and predictions about the future influence of AI. In addition, we decided to inquire about the current social climate within sport studies in Germany in order to gain insight into the situation and to potentially facilitate socio-cultural comparisons in subsequent research efforts. To this end, we included Jänsch and Bosse’s (2018) scale for measuring study requirements in the questionnaire.

*Online questionnaire*

The final set of 58 items resulted in the final sub-dimensions: “Usage Behavior” (10 items), “Motivation” (20 items), “Output Quality” (7 items), “Competencies” (6 items), “Future Ambitions and Expectations” (7 items), and “General Awareness” (8 items). The questionnaire was then migrated by two independent researchers from our team to “SoSci Survey”, an online platform designed for non-commercial scientific research. The team then spent several days reviewing the questionnaire.

*Pretest*

The pretesting phase consisted of two stages. First, scientists involved and not involved in the development of the questionnaire were invited to complete the questionnaire and provide feedback, especially on statistical or comprehension issues that arose. Subsequently, ten sports students from the University of Göttingen and all researchers from the Institute of Sport Science were asked to complete the questionnaire and provide feedback. This approach allowed us to include feedback from researchers and students outside the field of sport sociology. The feedback was analyzed by two researchers from the project team. Subsequently, a technical issue was addressed and an adjustment was made to a response scale inquiring about the frequency of AI use.

*Data collection*

Data collection took place between August 28, 2023 and November 30, 2023. To facilitate this process, a comprehensive list of all sports institutes in Germany was compiled and appropriate contact persons were identified. They received a description of the study along with a link to the survey. In addition, relevant communities, such as the German Society of Sport Science (dvs), were contacted via appropriate email distribution lists and asked to distribute the questionnaire to sports students. A total of 62 institutions were contacted, although not all were able to distribute our questionnaire, primarily due to internal university regulations and data protection policies.

*Statistical analyses*

The goal of this study was to provide initial empirical insights into a rapidly growing phenomenon and to establish an empirical framework for future in-depth research. The quantitative data were analyzed primarily using descriptive statistics. The following were examined: demographics, students' usage patterns, motivations for using AI, factors influencing students' motivation, uncertainties surrounding AI use, and their future perspectives on AI use.
Differences in mean values for specific items were examined to compare the AI variables presented. Furthermore, multivariate analysis was used to evaluate the relationship between socio-demographic variables, such as study progress, and AI-specific items. Missing cases were excluded listwise unless otherwise specified. Differences in the total number of cases resulting from questionnaire filters set at the beginning of the questionnaire were reported for each variable.

*Research Ethics and informed consent*

Our study is based on the ethical guidelines of the Georg-August-University. In addition, we contacted the university's data protection officer to ensure that the questionnaire met ethical standards. As a result of this consultation, certain sections and specific items were revised or removed from the initial version of our questionnaire. In addition, two documents containing detailed information on data protection were provided to the participants before the start of the study. Informed consent was also obtained from the participants. Students participated in our study on a voluntary basis and received no financial compensation for their participation. They were also informed of their right to withdraw at any time without consequence. In addition, the transcribed qualitative data (from a free-text field) are stored on two external storage devices at the Georg-August-Universität Göttingen.

# 5
# Sample

In our study, a total of 262 students participated; 39 of them discontinued the questionnaire during the response process. Among the participants, 118 identified as female, 101 as male, 2 as diverse, and 5 did not provide a response to this question. The participants had a mean age of 24.16 years (*SD* = 3.97), with ages ranging from 18 to 54 years.

Table 1: Gender

| **Gender** | *Frequency* | *Frequency percent* |
|---|---|---|
| Male | 101 | 38.5 |
| Female | 118 | 45.0 |
| Diverse | 2 | 0.8 |
| Missing (incl. no answer) | 41 | 15.6 |

"Missing (incl. no answer)" combines participants who did not reach the question (system-missing) with item nonresponse (no-answer codes); n = 41 (15.6%).

In terms of nationality, 83.6% of participants stated that they were German, with one person each from Denmark, Austria, Russia, and Vanuatu, whereas 39 did not specify their nationality.

Table 2: Nationality and migration backgrounds

| **Nationality** | *Frequency* | *Frequency percent* |
|---|---|---|
| Germany | 219 | 83.6 |
| Denmark | 1 | 0.4 |
| Austria | 1 | 0.4 |
| Russia | 1 | 0.4 |
| Vanuatu | 1 | 0.4 |
| Missing values | 39 | 14.9 |
| **Migration Background** | | |
| Yes | 13 | 5.0 |
| No | 208 | 79.4 |
| Not knowing | 1 | 0.4 |
| No response | 1 | 0.8 |
| Missing | 39 | 14.9 |

46.9% of the students reported being "single", 23.3% "living together", and 9.9% "living separately". Moreover, 41.6% reported living "with friends", 15.3% "with their partner", 14.5% "alone", and 11.8% "with relatives".

Table 3: Living situation

| **Family Status** | *Frequency* | *Frequency percent* |
|---:|:---:|:---:|
| Single | 123 | 46.9 |
| Together living[1] | 61 | 23.3 |
| Separate living[1] | 26 | 9.9 |
| Divorced | 1 | 0.4 |
| Widowed | 1 | 0.4 |
| Missing values | 50 | 19.1 |
| **Children** | | |
| Yes | 4 | 1.5 |
| No | 215 | 82.1 |
| Missing values | 43 | 16.4 |
| **Living situation** | | |
| With friends | 109 | 41.6 |
| With partner | 40 | 15.3 |
| With relatives | 31 | 11.8 |
| Alone | 38 | 14.5 |
| Nursing home | 1 | 0.4 |
| Not knowing | 2 | 0.8 |
| Missing | 41 | 15.6 |

[1] Married or in partnership

To better understand the students' backgrounds and gain insights into their lifestyles, we also inquired about their engagement in sports. As expected, the majority of the sports students are highly active. Specifically, 56.5% reported engaging in more than 4 hours of sports per week, while 26.4% stated they practice sports for 1-4 hours per week. Three participants reported being less active, engaging in sports for less than 1 hour per week, and one person indicated no involvement in sports at all.

Table 4: Sport behavior

| **Sport behavior** | *Frequency* | *Frequency percent* |
|---:|:---:|:---:|
| More than 4h week | 148 | 56.5 |
| 1-4h per week | 69 | 26.4 |
| Less than 1h per week | 3 | 1.1 |
| No sporting activity | 1 | 0.4 |
| Missing | 41 | 15.6 |

The academic progress of the sports students is quite balanced. The majority of the sports students are between the 3rd and 4th semesters. Very few are in their 13th semester or higher.

Table 5: Academic progress

| **Academic progress** | *Frequency* | *Frequency percent* |
|---|---|---|
| 1.-2. Semester | 28 | 10.7 |
| 3.-4. Semester | 47 | 17.9 |
| 5.-6. Semester | 38 | 14.5 |
| 7.-8. Semester | 40 | 15.3 |
| 9.-10. Semester | 31 | 11.8 |
| 11.-12. Semester | 23 | 8.8 |
| 13. Semester or higher | 14 | 5.3 |
| Missing | 41 | 15.6 |

We also asked for the educational level of the students, with multiple answers possible. The majority of the students are still studying. 85 have already received a Bachelor's degree, and 6 have a Master's degree. 13 stated they do not have any degree and are not currently in education.

Table 6: Educational status

| **Educational status** | *Frequency* | *Frequency percent* |
|---|---|---|
| Still in occupational education | 98 | 42.4 |
| No degree and not in education | 13 | 5.6 |
| Apprenticeship completed | 11 | 4.8 |
| Degree vocational school | 5 | 2.2 |
| School-Degree in healthcare (1yr) | 1 | 0.4 |
| School-Degree in healthcare (2-3 yrs) | 3 | 1.3 |
| Degree as educator | 1 | 0.4 |
| Degree of college of the GDR | 1 | 0.4 |
| Degree of tech. / admin. school | 2 | 0.9 |
| Bachelor | 85 | 36.8 |
| Diploma | 1 | 0.4 |
| Master | 6 | 2.6 |
| PhD | 1 | 0.4 |
| Other | 3 | 1.3 |

# 6

# Empirical results

Before reporting the descriptive findings, we examined the internal consistency of the multi-item scales used in this section. Because several of the AI-related blocks combined benefit- and concern-related items, negatively worded items were reverse-coded prior to analysis. Table 6a reports, for each scale, the number of items, the number of valid (listwise) cases, the mean, the standard deviation, and Cronbach's α.

Table 6a: Internal consistency and descriptive statistics of the multi-item scales

| **Scale** | **Items** | ***n*** | ***M*** | ***SD*** | **Cronbach's α** |
|---|---|---|---|---|---|
| Attitudes toward AI[1] (KI01) | 20 | 92 | 3.39 | 0.44 | .75 |
| Perceived output quality[1] (KI02) | 7 | 97 | 2.43 | 0.45 | .54 |
| AI competencies (KI03) | 6 | 90 | 3.29 | 0.57 | .58 |
| Future expectations[1] (KI04) | 7 | 203 | 3.28 | 0.60 | .64 |
| General appraisal of AI[1] (KI05) | 8 | 181 | 3.30 | 0.50 | .56 |
| Study demands / social climate (SA01) | 21 | 163 | 3.42 | 0.51 | .86 |

Responses were given on a five-point Likert-type scale; higher values indicate stronger agreement with the respective item. n = listwise complete cases; M and SD refer to the scale mean. [1] Negatively worded items were reverse-coded prior to analysis. The study-demands (social climate) scale was adapted from Jänsch and Bosse (2018).

In exploratory subgroup comparisons (Mann–Whitney U tests; effect sizes reported as rank-biserial correlations, r), male students reported higher self-rated AI competence than female students ($M = 3.50$ vs. 3.11; $p = .001$, $r = .35$), with smaller gender differences for attitudes toward AI ($p = .045$, $r = .22$) and future expectations ($p = .014$, $r = .19$). Students who reported having already used AI tools showed somewhat more positive future expectations ($p = .041$, $r = .16$) and general appraisals ($p = .015$, $r = .20$) than those who had not. Across scales, attitudes toward AI correlated moderately with perceived competence (Spearman's $\rho = .45$) and future expectations ($\rho = .49$), whereas perceived output quality was largely unrelated to the other scales ($\rho \leq .18$). Owing to the self-selected convenience sample and the heterogeneity of some scales, these findings are exploratory and should be read as hypothesis-generating rather than confirmatory.

## 6.1 Use of AI tools

In this section, we present empirical findings on the usage behavior and interest in using AI tools among sports students in their studies. In addition, we investigated the students' familiarity with various AI tools and whether they were aware of university courses on AI or had even participated in such courses. Two-thirds of the students said they had already used AI tools in their studies.

Table 7: Number of students who have ever used AI

| **Have you ever used AI-tools?** | *response* | *n* | *percent* |
|---|---|---|---|
| Total | Yes | 168 | 67.5 |
| | No | 81 | 32.5 |

Among them, slightly more men (71.3%) than women (62.7%) said that they used AI tools.

Table 8: Number of students who have ever used AI, separated by gender [1]

| **Have you ever used AI tools?** | *response* | *n* | *percent among genders* |
|---|---|---|---|
| Female | Yes | 74 | 62.7 |
| | No | 44 | 37.3 |
| Male | Yes | 72 | 71.3 |
| | No | 29 | 28.7 |
| Diverse | Yes | 1 | 50.0 |
| | No | 1 | 50.0 |
| No gender stated | Yes | 2 | 50.0 |
| | No | 2 | 50.0 |
| Actively “not stated” | | 1 | |

[1] The difference in the total number of responses between Table 7 and 8 results can be explained by the fact that not all students who replied to that question also stated their gender.

Looking at the absolute frequency of students who have ever used AI, 74 identified as female, 72 identified as male, and 1 identified as diverse. The group of students who have never used AI tools is divided into 44 females, 29 males, and 1 diverse student.

Figure 1: Use of AI, separated by Gender (absolute figures)

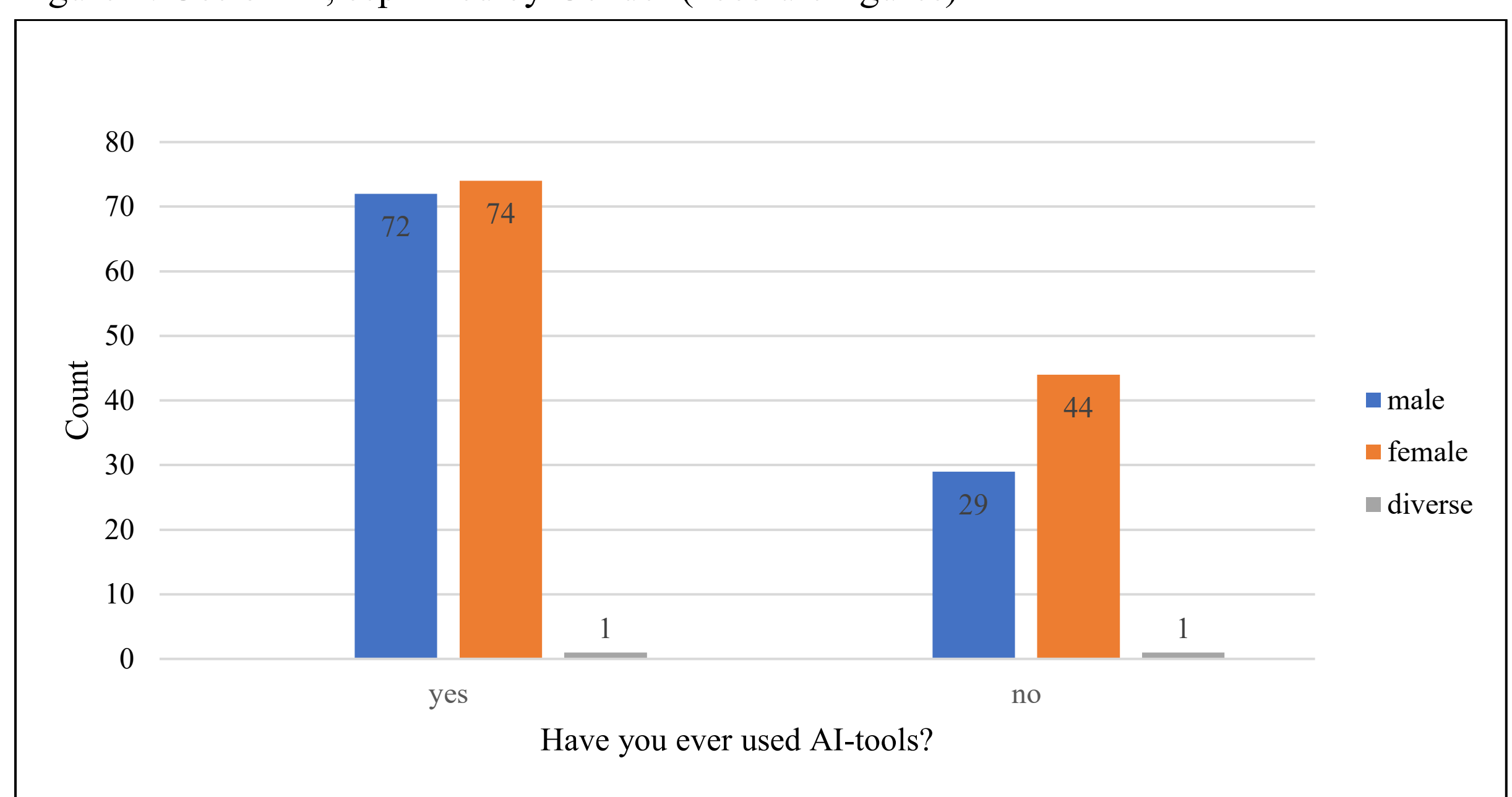

In terms of student progression, the highest rate (82.6%) of sport students who had ever used AI tools was between the 11th and 12th semester. The lowest rate (54.8%) of having ever used AI occurred among students with a study progress of 9th to 10th semester. It should be noted that the total number of sport students for each progress option varies.

Table 9: Students who have ever used AI, separated by academic progress

| **Have you ever used AI tools?** | *response* | *n* | *percent among semesters* |
|---|---|---|---|
| 1.-2. Semester | Yes | 17 | 60.7 |
| | No | 11 | 39.3 |
| 3.-4. Semester | Yes | 35 | 74.5 |
| | No | 12 | 25.5 |
| 5.-6. Semester | Yes | 24 | 63.2 |
| | No | 14 | 36.8 |
| 7.-8. Semester | Yes | 26 | 65.0 |
| | No | 14 | 35.0 |
| 9.-10. Semester | Yes | 17 | 54.8 |
| | No | 14 | 45.2 |
| 11.-12. Semester | Yes | 19 | 82.6 |
| | No | 4 | 17.4 |
| 13. Semester or higher | Yes<br>No | 10<br>4 | 71.4<br>28.6 |
| Actively "not stated" | | 2 | |

To assess students' knowledge of the existence of different AI tools, we provided a list of common applications. Most students have heard of ChatGPT with 98%, followed by Bing with 15.7% and Google Bard with 14.5%.

Table 10: Known AI tools[1]

| **Known AI tools** | *percent known AI tools* |
|---|---|
| ChatGPT | 98.0 |
| Bing Chat (now MS-Copilot) | 15.7 |
| Google Bard (now Google Gemini) | 14.5 |
| Midjourney | 8.8 |
| Notion AI | 7.2 |
| CopyAI | 6.0 |
| GitHub Copilot | 5.2 |
| Jasper | 3.6 |
| Lightning AI | 3.6 |
| Slides AI | 3.2 |
| Krisp | 2.8 |
| Albus | 2.4 |
| YouChat | 2.4 |
| RocketAI | 2.0 |
| Voicemod | 2.0 |
| None | 1.2 |
| Runway | 0.8 |
| Altered | 0.4 |
| Flikl | 0.4 |
| SuperCreator | 0.4 |
| Other | 9.2 |
| None | 1.2 |

[1] Multiple answers possible. Percentages in the right column are based on the 249 participants who answered this question; they sum to more than 100% because more than one answer could be selected.

We also asked about the frequency with which students use AI tools in their studies. 14.9% of them reported that they use AI tools several times a week, 11.3% once a week, and hardly anyone uses AI tools every day (0.6%). 22.0% use AI tools one to three days per month. The majority of students (33.3%) said they use AI tools less than once to three times a month. 17.9% stated they have never used AI tools in their sports studies.

Table 11: Frequency of use of AI tools[1]

| **General use of AI tools** | *n* | *percent* |
|---|---|---|
| Every day | 1 | 0.6 |
| Several times per week | 25 | 14.9 |
| Once a week | 19 | 11.3 |
| One to three times per month | 37 | 22.0 |
| Less than one to three times per month | 56 | 33.3 |
| Never | 30 | 17.9 |

[1] 168 out of 262 students responded to that question. The missing 94 cases consist of 81 participants who responded “no” to the question “Have you ever used AI tools?” at the beginning of the questionnaire and 13 who did not respond to this specific question. A further 30 respondents selected “Never” here, indicating that they have not used AI tools in their studies but have used them in other contexts.

Students were also asked which specific AI tools they use in their studies. The majority (62.6%) reported using ChatGPT, followed by Bing Chat (3.2%) and Notion AI (2.6%). A further 22.1% indicated that they did not use any of the listed tools, while 4.2% named additional tools in a free-text field.

Table 12: Frequency of use of AI tools, separated by specific tools

| **Use of specific AI tools** | *n* | *percent* |
|---|---|---|
| ChatGPT | 119 | 62.6 |
| Bing Chat | 6 | 3.2 |
| Notion AI | 5 | 2.6 |
| Google Bard | 4 | 2.1 |
| Midjourney | 2 | 1.1 |
| CopyAI | 1 | 0.5 |
| GitHub Copilot | 1 | 0.5 |
| Slides AI | 1 | 0.5 |
| YouChat | 1 | 0.5 |
| Other[1] | 8 | 4.2 |
| None | 42 | 22.1 |

[1] The additional AI tools mentioned in the free text field are: Adobefly, AlephAlpha, Alpaca, AlphaFold, AlphaZero, Buzzwrite, Canva, Canva text to picture, Deepl, DeepL Write, DeppGPT by Postillon, Elicit, Goat Chat, GPT, GPT-2, GPT-3, Grammarly Jenni AI, LambdaZero, Languagetool, Llama by Facebook, PDFgear, Snapchat, Stable diffusion, Steve.ai, Upscale img, Writefull, and WriteSonic.

Furthermore, we inquired as to whether students were aware of their university's provision of courses designed to enhance the handling of AI tools. Only 11.1% of respondents indicated that their university did indeed offer such courses. The majority, comprising 64.9%, indicated that they were unsure, while 19.1% stated that their universities did not offer any courses on AI.

Table 13: Information about university courses on AI

| **Information about university courses on AI** | *n* | *percent* |
|---:|:---:|:---:|
| Yes | 29 | 11.1 |
| No | 50 | 19.1 |
| I don't know | 170 | 64.9 |
| Missing | 13 | 5 |

In particular, students in their first semester were uncertain as to whether their university offers such courses. Upon examination of the responses across individual semesters, with the exception of the first two semesters, the majority of students indicated that there were no AI courses at their university.

Table 14: Information about university courses on AI, separated by semesters

| **Information about university courses on AI, separated by semesters** | *response* | *n* | *percent among semesters* |
|---:|:---:|:---:|:---:|
| 1.-2. Semester | Yes | 4 | 14.3 |
| | No | 2 | 7.1 |
| | Don't know | 22 | 78.6 |
| 3.-4. Semester | Yes | 1 | 2.1 |
| | No | 6 | 12.8 |
| | Don't know | 40 | 85.1 |
| 5.-6. Semester | Yes | 5 | 13.2 |
| | No | 10 | 26.3 |
| | Don't know | 23 | 60.5 |
| 7.-8. Semester | Yes | 5 | 12.5 |
| | No | 10 | 25.0 |
| | Don't know | 25 | 62.5 |
| 9.-10. Semester | Yes | 6 | 19.4 |
| | No | 7 | 22.6 |
| | Don't know | 18 | 58.1 |
| 11.-12. Semester | Yes | 3 | 13.0 |
| | No | 4 | 17.4 |
| | Don't know | 16 | 69.6 |
| 13. Semester or higher | Yes | 3 | 21.4 |
| | No | 4 | 28.6 |
| | Don't know | 7 | 50.0 |
| "No information" | Don't know | 2 | 100 |

At the time of the survey, only 4% of the students had participated in university courses dealing with AI, whereas 96% had never attended those courses or answered indecisively.

Table 15: University courses attended

| **University courses on AI attended** | *n* | *percent* |
|---|---|---|
| yes | 10 | 3.8 |
| no | 237 | 90.5 |
| Don't know | 2 | .8 |
| Missing | 13 | 5.0 |

A total of 75 students who were in the early stages of their studies did not attend any AI courses at their universities. Even in the higher semesters, the number of participants who had already attended such courses during their study phase remained low.

Table 16: University courses attended, separated by semesters

| **University courses on AI attended, separated by semesters** | *yes* | *no* | *don't know* |
|---|---|---|---|
| 1.-2. Semester | 0 | 28 | 0 |
| 3.-4. Semester | 0 | 47 | 0 |
| 5.-6. Semester | 1 | 37 | 0 |
| 7.-8. Semester | 2 | 38 | 0 |
| 9.-10. Semester | 3 | 28 | 0 |
| 11.-12. Semester | 1 | 22 | 0 |
| 13. Semester or higher | 2 | 11 | 1 |
| Total | 9 | 211 | 1 |

As demonstrated in Table 17, the calculated contingency measure regarding the hypothesized relationship between study progress and the attendance of university courses is 0.32, indicating a low connection. With a significance value of 0.016, it can be inferred that there is a significant correlation between these two variables. The effect size can be assessed as moderate (*Phi* = 0.34, *Cramer-V* = 0.24).

Table 17: Contingency and effect size study progress and attendance of courses

| **Symmetrical measures** | | *value* | *Approximately significance* |
|---|---|---|---|
| Nominal by nominal measure | Phi | .335 | .016 |
| | Cramer-V | .237 | .016 |
| | Coefficient for contingency | .318 | .016 |

## 6.2 Motivational factors for the use of AI tools among students

This section presents empirical results on the motivational factors of sports students in relation to the use of AI tools.

Figure 2: Application areas of AI tools

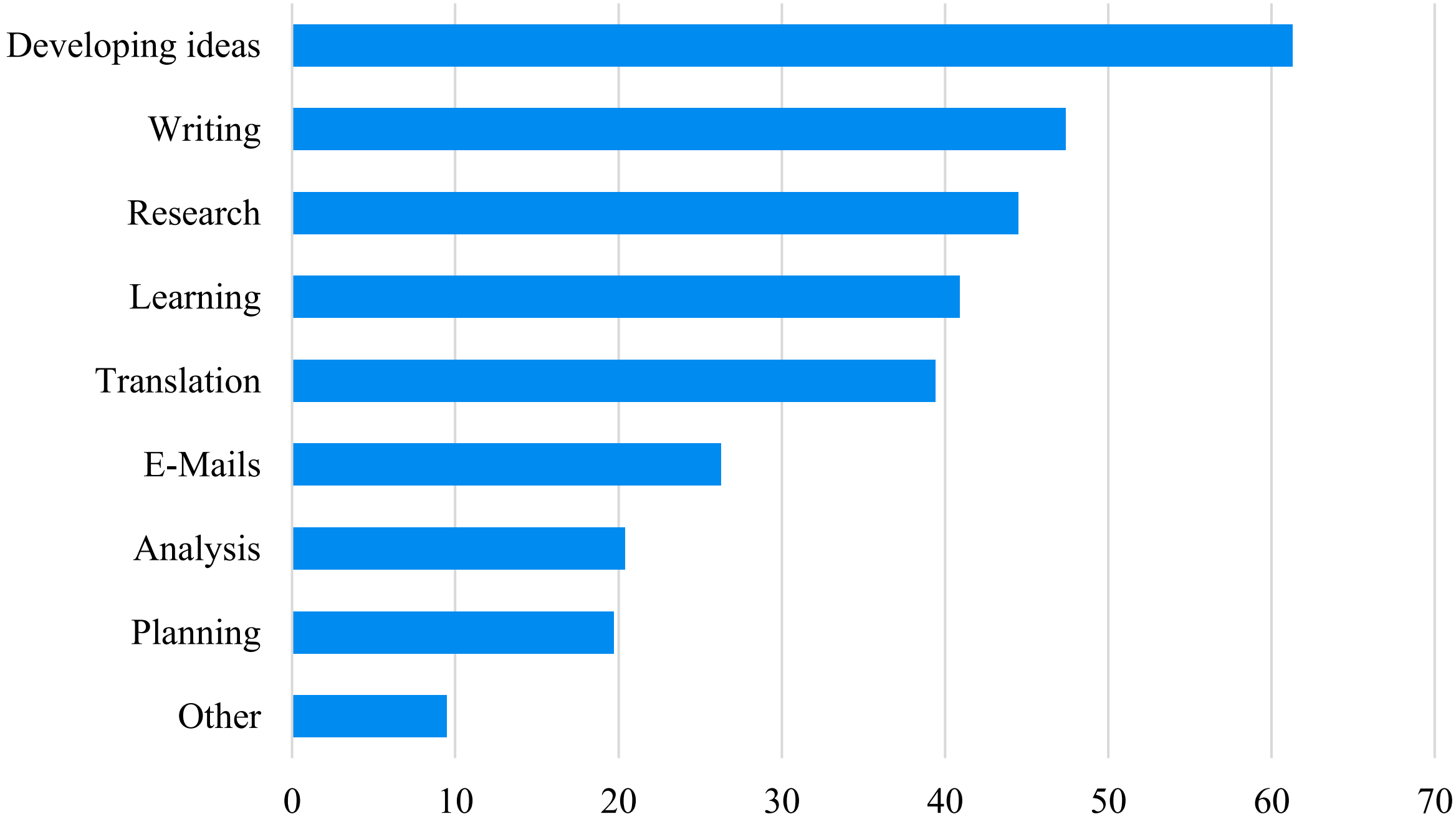

Data are presented in percent; multiple answers possible; valid set $n$ = 424 responses.
**Developing ideas** was specified with "for example for a research project"; **writing** "f. e. an essay"; **research**, "f. e. finding literature"; **learning**, "f. e. help understanding concepts"; **translation**, "f. e. from English to German"; **E-Mails**, "f. e. phrasing messages; **analysis**, "f. e. data calculation"; **planning**, "f. e. creating timetables".

The majority of students utilized AI tools primarily for idea development, such as for research projects (61.3%), followed by those who employed them for writing purposes, such as academic papers (47.4%). 44.5% reported using AI tools for research-related tasks, such as finding suitable literature, while 40.9% stated using them for learning purposes, such as understanding theoretical concepts or the viewpoints of relevant authors.

Table 18: Application areas of AI tools

| **Application areas of AI tools** | *n* | *percent* |
|---:|:---:|:---:|
| Developing ideas | 84 | 61.3 |
| Writing | 65 | 47.4 |
| Research | 61 | 44.5 |
| Learning | 56 | 40.9 |
| Translation | 54 | 39.4 |

| E-Mails | 36 | 26.3 |
|---|---|---|
| Analysis | 28 | 20.4 |
| Planning | 27 | 19.7 |
| Other | 13 | 9.5 |
| Total | 424 | 309.5 |

The other areas of potential usage mentioned by the students are as follows:

- Creating images for presentations
- Developing presentations
- Explaining concepts
- Providing faster responses compared to searching on Google or asking friends
- Providing brief explanations of particular terms
- Experimenting with new ideas
- Paraphrasing
- Fostering independent work
- Proofreading
- Addressing statistical questions
- Summarizing complex texts
- Finding synonyms
- Preparing job applications and resumes

The mean for the item "developing ideas" was 4.2 (*SD* = 0.78), indicating the highest level of motivation among sport students. In contrast, the item "planning" had the lowest mean of 3.7 (*SD* = 1.27). However, all of the mean scores were above the no response option, indicating that students tend to agree with the motivational potential of using AI tools in these areas.

Figure 3: Motivation of using AI tools, separated by specific areas

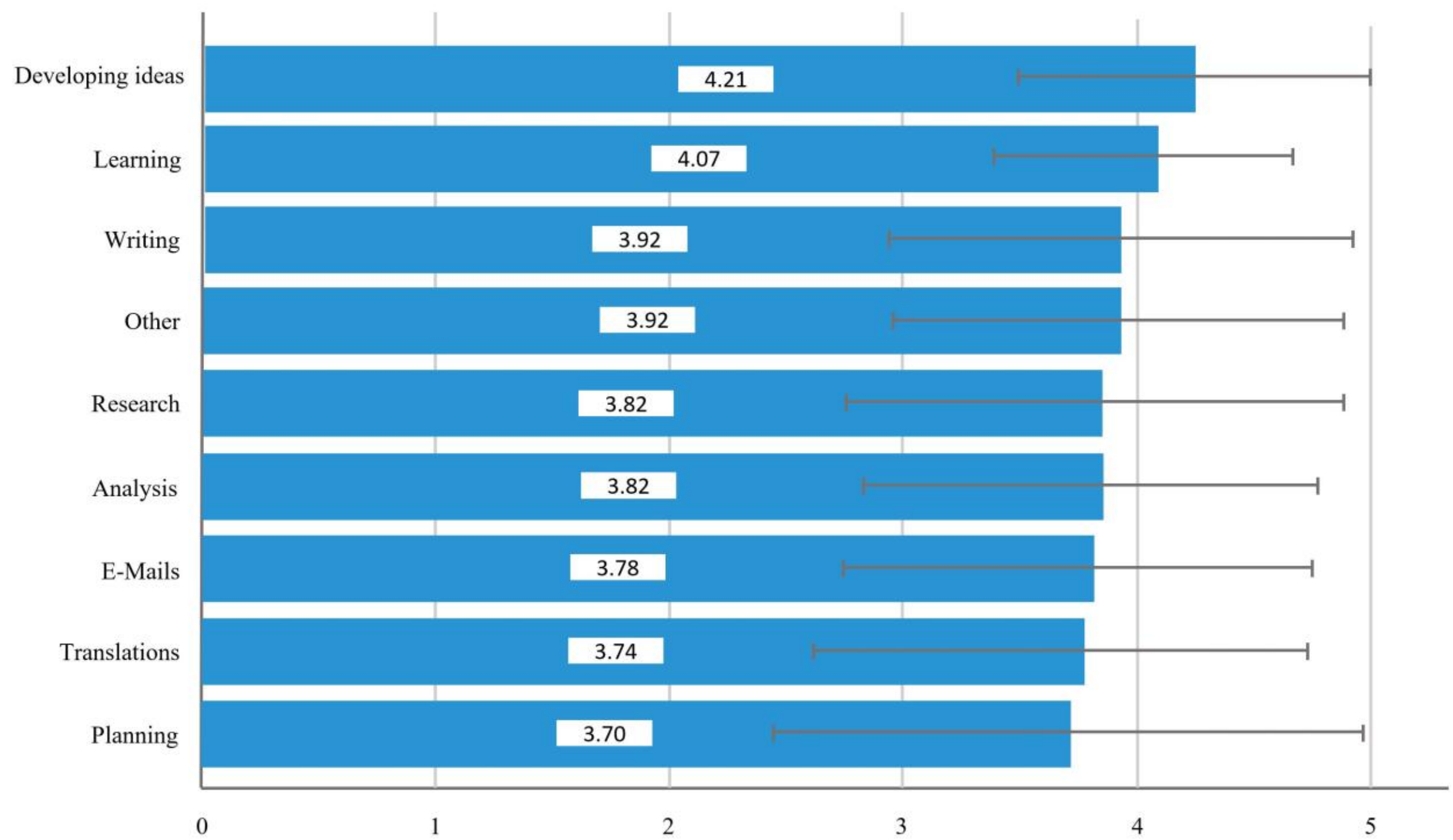

Responses were provided on a five-point Likert scale (5: strongly agree, 4: agree, 3: neither agree nor disagree, 2: disagree, 1: strongly disagree; or no response) following the item stem "The use of AI tools fosters my motivation in the following areas..."

When looking at the extreme values, there are small differences in the students' responses. With the exception of learning, e-mail, analysis, and others, which received at least a 2 (disagree), each area of use received at least the strongest disagreement (1).

Table 19: Motivation for the use of AI tools in specific areas

| **Motivation of AI use in specific areas** | *n* | *min* | *max* | *mean* | *sd* |
|---|---|---|---|---|---|
| Developing ideas | 84 | 1 | 5 | 4.21 | .78 |
| Learning | 56 | 2 | 5 | 4.07 | .66 |
| Writing | 64 | 1 | 5 | 3.92 | .98 |
| Other | 13 | 2 | 5 | 3.92 | .95 |
| Research | 61 | 1 | 5 | 3.82 | 1.04 |
| Analysis | 28 | 2 | 5 | 3.82 | .98 |
| E-Mails | 36 | 2 | 5 | 3.78 | 1.02 |
| Translations | 54 | 1 | 5 | 3.74 | 1.05 |
| Planning | 27 | 1 | 5 | 3.70 | 1.27 |

Students were also asked about their overall motivation to use AI tools. The results are shown in Figure 4. On average, the role of AI in practical sport education was rated as the least motivating factor ($M$ = 1.87, $SD$ = 1.12), while AI as supplementary learning source ($M$ = 4.03, $SD$ = 0.91) and saving time with AI ($M$ = 4.02, $SD$ = 0.93) received the highest approval.

Figure 4: Overall motivation among students to use AI tools

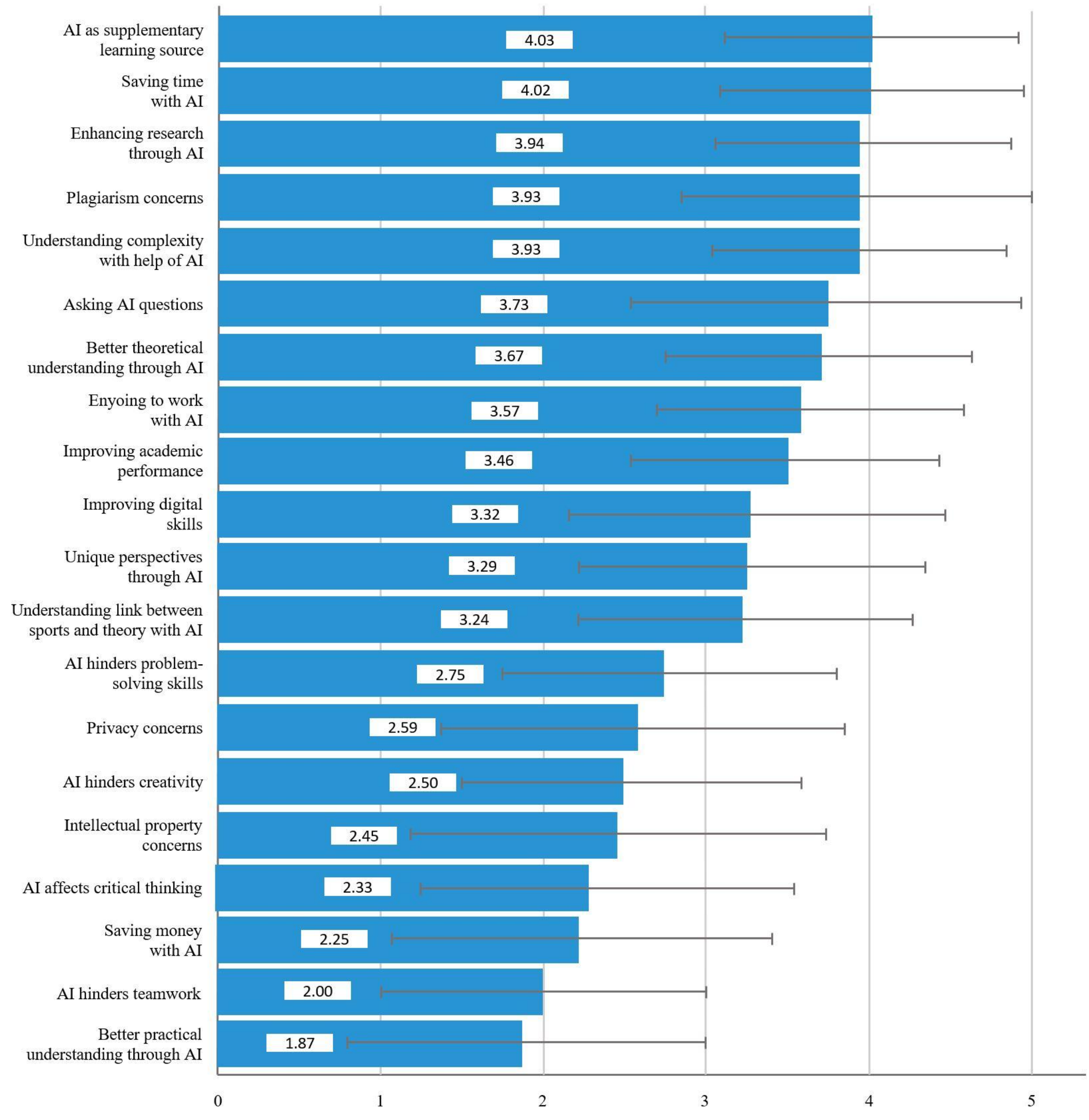

Answers were provided on a five-point Likert scale (1: strongly disagree, 2: disagree, 3: neither agree nor disagree, 4: agree, 5: strongly agree; or no response). The participants were instructed to indicate the extent to which they agree with the statements provided regarding the motivation for using AI tools in their sports studies.

Students' agreement with various motivational factors regarding the use of AI tools in their studies generally decreases when it comes to critical aspects. Except for worrying about plagiarism, which was rated fourth highest ($M = 3.93$, $SD = 1.09$).

Table 20: Overall motivation among students to use AI tools

| **Overall motivation to use AI tools** | *n* | *min* | *max* | *mean* | *sd* |
|---|---|---|---|---|---|
| AI as supplementary learning source | 127 | 1 | 5 | 4.03 | .91 |
| Saving time with AI | 126 | 1 | 5 | 4.02 | .93 |
| Enhancing research through AI | 127 | 1 | 5 | 3.94 | .89 |
| Plagiarism concerns | 125 | 1 | 5 | 3.93 | 1.09 |
| Understanding complexity with help of AI | 126 | 1 | 5 | 3.93 | .90 |
| Asking AI questions | 119 | 1 | 5 | 3.73 | 1.22 |
| Theoretical understanding with AI | 124 | 1 | 5 | 3.67 | 1.01 |
| Enjoying AI | 122 | 1 | 5 | 3.57 | .95 |
| Improving academic performance | 125 | 1 | 5 | 3.46 | .96 |
| Improving digital skills | 127 | 1 | 5 | 3.32 | 1.13 |
| Unique perspectives through AI | 124 | 1 | 5 | 3.29 | 1.02 |
| Understanding link between sports and theory with AI | 122 | 1 | 5 | 3.24 | 1.02 |
| AI hinders problem-solving skills | 122 | 1 | 5 | 2.75 | 1.03 |
| Privacy concerns | 123 | 1 | 5 | 2.59 | 1.25 |
| AI hinders creativity | 125 | 1 | 5 | 2.50 | 1.04 |
| Intellectual property concerns | 122 | 1 | 5 | 2.45 | 1.26 |
| AI affects critical thinking | 126 | 1 | 5 | 2.33 | 1.05 |
| Saving money with AI | 116 | 1 | 5 | 2.25 | 1.19 |
| AI hinders teamwork | 121 | 1 | 5 | 2.00 | .99 |
| Better practical understanding through AI | 119 | 1 | 5 | 1.87 | 1.12 |
| Valid N (listwise) | 92 | | | | |

### 6.3 AI competencies and AI output quality

Students who reported using AI in their sport studies were also asked to rate their own competence and the perceived competence of their lecturers in using AI tools. The highest level of agreement was found for items suggesting that lecturers (*M* = 4.16, *SD* = 0.8) and students (*M* = 4.13, *SD* = 0.85) should be trained in the use of AI tools. The lowest level of agreement was found for whether lecturers are adequately trained in AI (*M* = 1.96, *SD* = 0.78).

Figure 5: Competencies in using AI

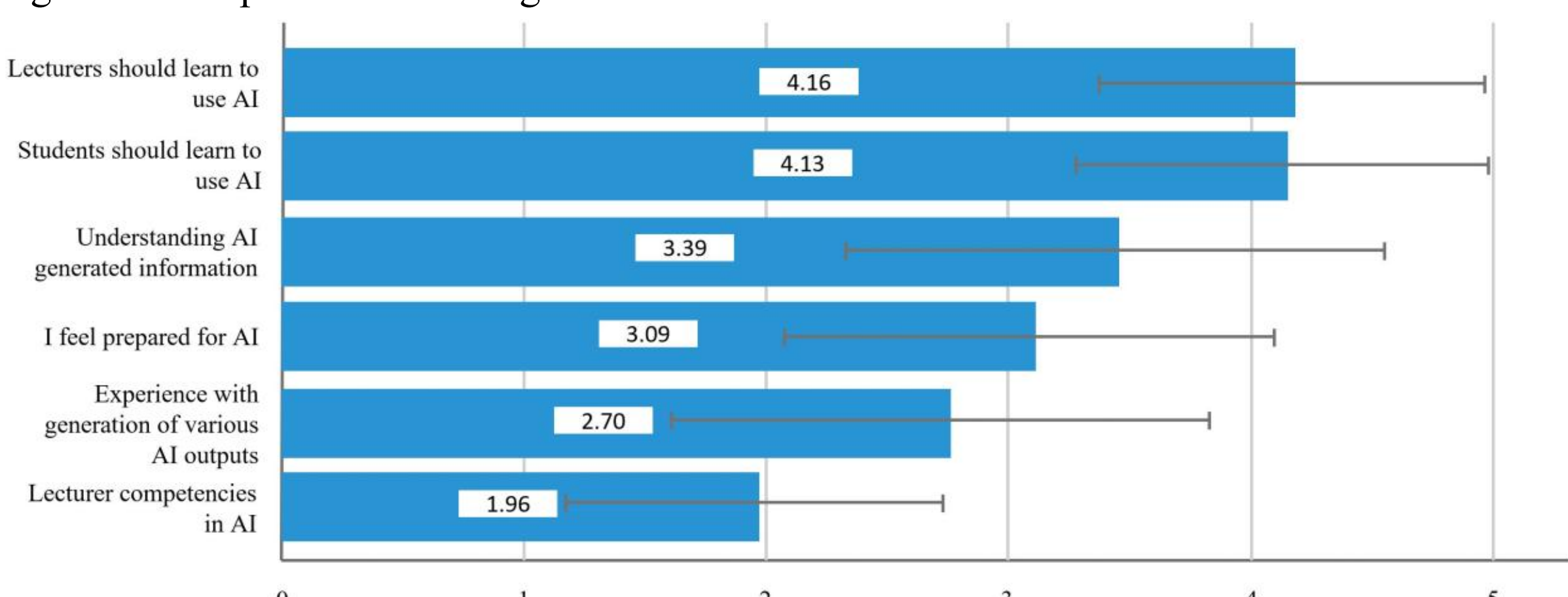

Responses were provided on a five-point Likert scale (1: strongly disagree, 2: disagree, 3: neither agree nor disagree, 4: agree, 5: strongly agree; or no response). The participants were instructed to indicate the extent to which they agree with the statements provided regarding the role of competency in AI tools in their sports studies.

Other competency items were rated with values around the middle: Understanding information generated by AI (*M* = 3.39, *SD* = 1.13), Feeling prepared for AI (*M* = 3.09, *SD* = 1.01), and Experience generating various AI outputs (*M* = 2.70, *SD* = 1.13).

Table 21: Competencies in using AI

| **Competencies in using AI** | *n* | *min* | *max* | *mean* | *sd* |
|---|---|---|---|---|---|
| Lecturers should learn to use AI | 119 | 1 | 5 | 4.16 | .80 |
| Students should learn to use AI | 121 | 1 | 5 | 4.13 | .85 |
| Understanding AI generated information | 122 | 1 | 5 | 3.39 | 1.13 |
| I feel prepared for AI | 121 | 1 | 5 | 3.09 | 1.01 |
| Experience with generation of various AI outputs | 122 | 1 | 5 | 2.70 | 1.13 |
| Lecturer competencies in AI | 90 | 1 | 5 | 1.96 | .78 |
| Valid N (listwise) | 90 | | | | |

To gain a better understanding of how students perceive artificial knowledge production, we also asked them to rate the quality of AI-generated information. On average, students mostly agreed with the issues that AI tools can generate misinformation ($M = 4.59$) and that they need to verify the information provided by AI in order to use it for their studies ($M = 4.48$). The item with the lowest agreement was "I perceive AI tools as reliable" ($M = 3.10$).

Figure 6: Quality assessment of AI generated information

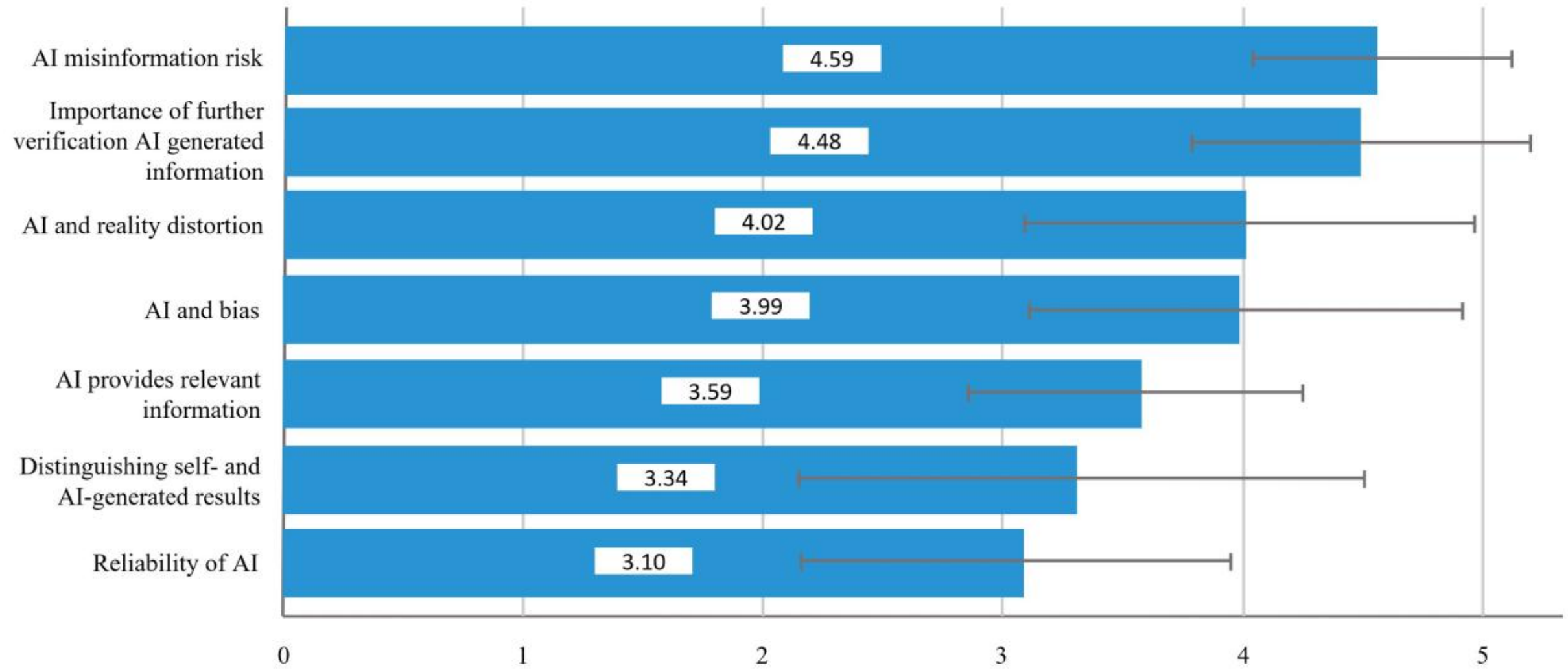

Responses were provided on a five-point Likert scale (1: strongly disagree, 2: disagree, 3: neither agree nor disagree, 4: agree, 5: strongly agree; or no response). The instructions requested that participants indicate their level of agreement with statements pertaining to the output quality of AI tools.

Table 22: Quality assessment of AI generated information

| **Quality of AI generated information** | *n* | *min* | *max* | *mean* | *sd* |
|---|---|---|---|---|---|
| AI misinformation risk | 123 | 3 | 5 | 4.59 | .54 |
| Importance of further verification of AI generated information | 124 | 2 | 5 | 4.48 | .74 |
| AI and reality distortion | 113 | 1 | 5 | 4.02 | .92 |
| AI and bias | 109 | 1 | 5 | 3.99 | .89 |
| AI provides relevant information | 126 | 1 | 5 | 3.59 | .75 |
| Distinguishing self- and AI-generated results | 116 | 1 | 5 | 3.34 | 1.13 |
| Reliability of AI | 124 | 1 | 5 | 3.10 | .84 |
| Valid N (listwise) | 97 | | | | |

### 6.4 Uncertainties and future expectations

In the following chapter, we explore students' perspectives on uncertainties and future expectations regarding the use of AI tools. In terms of motivational issues, when examining students' uncertainties about the use of AI tools in their studies, concerns about plagiarism received the most agreement (*M* = 3.93, *SD* = 1.10), while students tended to disagree with the idea that AI would hinder teamwork (*M* = 2.00, *SD* = 0.99), affect critical thinking (*M* = 2.33, *SD* = 1.05), or lead to intellectual property concerns (*M* = 2.45, *SD* = 1.26).

Figure 7: Uncertainties about the use of AI

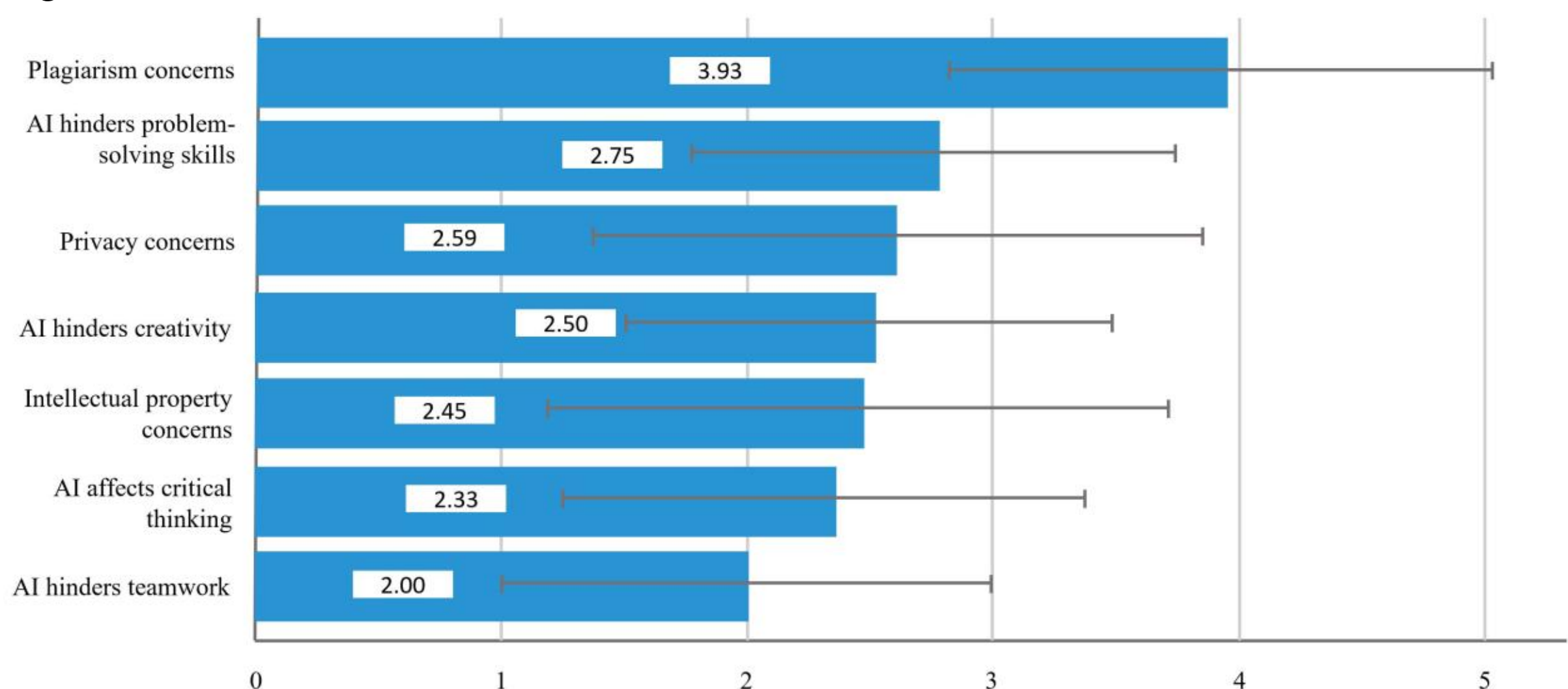

Responses were provided on a five-point Likert scale (1: strongly disagree, 2: disagree, 3: neither agree nor disagree, 4: agree, 5: strongly agree; or no response). The items presented are extracted from the motivational question field discussed in Chapter 6.3 in order to highlight issues that have been perceived as critical.

Table 23: Uncertainties about the use of AI

| **Uncertainties about the use of AI** | *n* | *min* | *max* | *mean* | *sd* |
|---|---|---|---|---|---|
| Plagiarism concerns | 125 | 1 | 5 | 3.93 | 1.10 |
| AI hinders problem-solving skills | 122 | 1 | 5 | 2.75 | 1.03 |
| Privacy concerns | 123 | 1 | 5 | 2.59 | 1.25 |
| AI hinders creativity | 125 | 1 | 5 | 2.50 | 1.04 |
| Intellectual property concerns | 122 | 1 | 5 | 2.45 | 1.26 |
| AI affects critical thinking | 126 | 1 | 5 | 2.33 | 1.05 |
| AI hinders teamwork | 121 | 1 | 5 | 2.00 | .99 |
| Valid N (listwise) | 113 | | | | |

More broadly, we also asked students to provide general feedback on the potential negative impacts of AI on academia. With a mean of *M* = 3.01, "Negative impact of AI on learning" (*SD*

= 1.12) received the highest agreement, followed by “AI threatens academic jobs” with a mean of *M* = 2.65 (*SD* = 1.12).

Figure 8: General uncertainties in relation to AI in academia

Responses were provided on a five-point Likert scale (1: strongly disagree, 2: disagree, 3: neither agree nor disagree, 4: agree, 5: strongly agree; or no response). The items presented are extracted from the questions regarding future perspectives for addressing critically perceived issues.

Table 24: General uncertainties in relation to AI in academia

| **Uncertainties in relation to AI in academia** | *n* | *min* | *max* | *mean* | *sd* |
|---|---|---|---|---|---|
| Negative impact of AI on learning | 222 | 1 | 5 | 3.01 | 1.12 |
| AI threatens academic jobs | 225 | 1 | 5 | 2.65 | 1.12 |
| AI reduces student interactions | 224 | 1 | 5 | 2.52 | 1.22 |
| Valid N (listwise) | 214 | | | | |

Additionally, students were asked to assess the future impact of AI on the society. The majority agreed with the unstoppable development (*M* = 4.19, *SD* = 0.8) and influence of AI (*M* = 4.09, *SD* = 0.77), followed by concerns about AI can cause societal crises (*M* = 3.42, *SD* = 1.0) and a general uncertainty about the future impact of AI (*M* = 3.25, *SD* = 1.17). Concerns about AI exacerbating social inequality (*M* = 2.91, *SD* = 1.11) and AI being a step toward surveillance (*M* = 2.65, *SD* = 1.04) received the lowest levels of support.

Figure 9: Future negative impact of AI on society

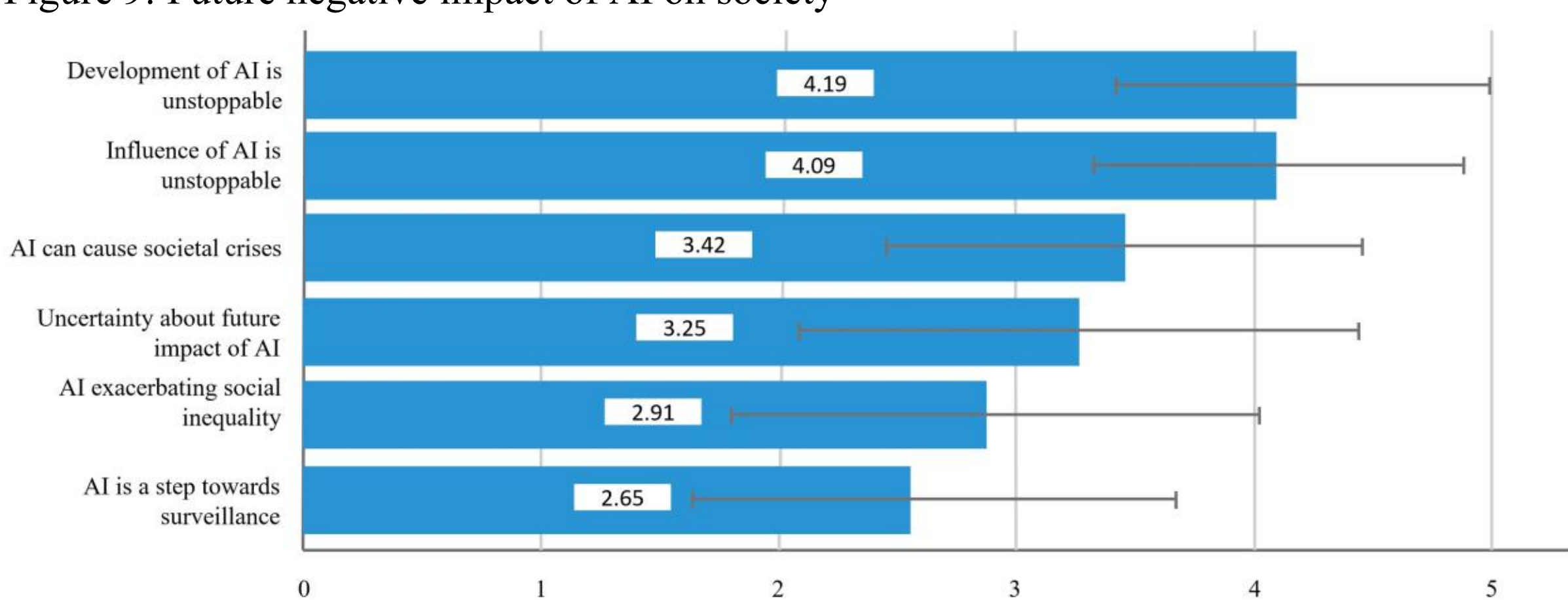

Responses were provided on a five-point Likert scale (1: strongly disagree, 2: disagree, 3: neither agree nor disagree, 4: agree, 5: strongly agree; or no response). The items presented were extracted from the set of questions on general perceptions of AI in order to present critically perceived issues.

Table 25: Future negative impact of AI on society

| **Future negative impact of AI on society** | *n* | *min* | *max* | *mean* | *sd* |
|---|---|---|---|---|---|
| Development of AI is unstoppable | 221 | 1 | 5 | 4.19 | .80 |
| Influence of AI is unstoppable | 222 | 1 | 5 | 4.09 | .77 |
| AI can cause societal crises | 201 | 1 | 5 | 3.42 | 1.00 |
| Uncertainty about future impact of AI | 221 | 1 | 5 | 3.25 | 1.17 |
| AI exacerbating social inequality | 207 | 1 | 5 | 2.91 | 1.11 |
| AI is a step towards surveillance | 215 | 1 | 5 | 2.65 | 1.04 |
| Valid N (listwise) | 191 | | | | |

At the end of the questionnaire, students were also asked about the potential positive effects of AI and its impact on their sport studies.

Figure 10: Future positive impact of AI on society and studies

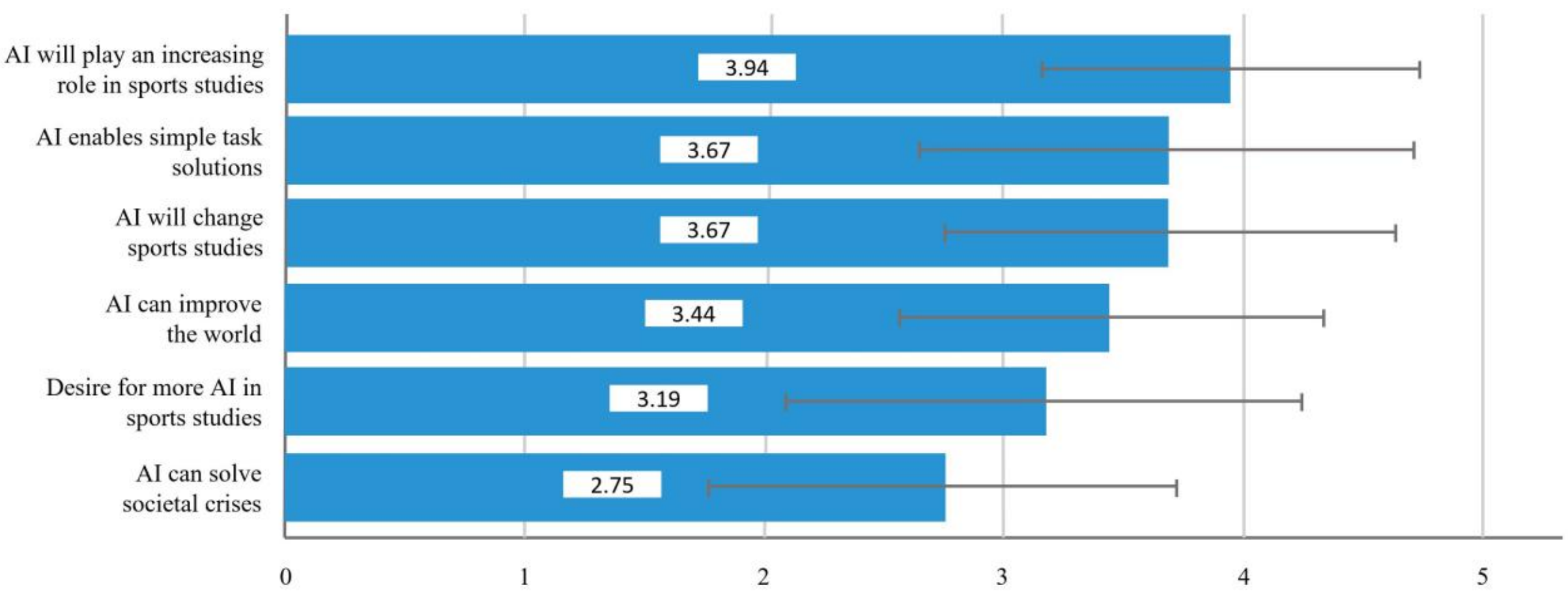

Responses were provided on a five-point Likert scale (1: strongly disagree, 2: disagree, 3: neither agree nor disagree, 4: agree, 5: strongly agree; or no response).

Sports students on average agree that AI tools will play an increasing role in their future studies ($M$ = 3.94, $SD$ = 0.77), while there is less consensus on the capabilities of AI to solve societal crises ($M$ = 2.74, $SD$ = 0.96).

Table 26: Future positive impact of AI on society and studies

| **Future positive impact of AI on society and studies** | *n* | *min* | *max* | *mean* | *sd* |
|---|---|---|---|---|---|
| AI will play an increasing role in sports studies | 225 | 1 | 5 | 3.94 | .77 |
| AI enables simple task solutions | 227 | 1 | 5 | 3.67 | .99 |
| AI will change sports studies | 224 | 1 | 5 | 3.67 | .94 |
| AI can improve the world | 210 | 1 | 5 | 3.44 | .90 |
| Desire for more AI in sports studies | 221 | 1 | 5 | 3.19 | 1.10 |
| AI can solve societal crises | 200 | 1 | 5 | 2.74 | .96 |
| Valid N (listwise) | 187 | | | | |

## 6.5 Social climate in sports studies in Germany

Sport students responded to the items of the validated scale by Jänsch and Bosse (2018) regarding their perception of study demands on a five-point Likert scale ranging from 1 (very difficult) to 5 (very easy), and were also given the option "does not fit for me." In terms of their overall study demands, students tend to perceive their study demands as slightly easier than harder, with a mean of $M = 3.43$ ($SD = 0.52$).

Item mean values are shown in Tab. 27 to Tab. 32 and Fig. 11 to Fig. 17. Means and error bars are presented for the following subscales: Contact and Cooperation, Study Organization, Performance Pressure, Learning Activity, Scientific Method, and Study Expectations. As can be seen in Figure 11, on the contact and cooperation subscale, the student-athletes rated their coping with the social climate as the easiest ($M = 4.02$, $SD = 0.86$) study demand, while the organization of teamwork reached the lowest mean value ($M = 3.42$, $SD = 1.14$).

Figure 11: Contact and cooperation

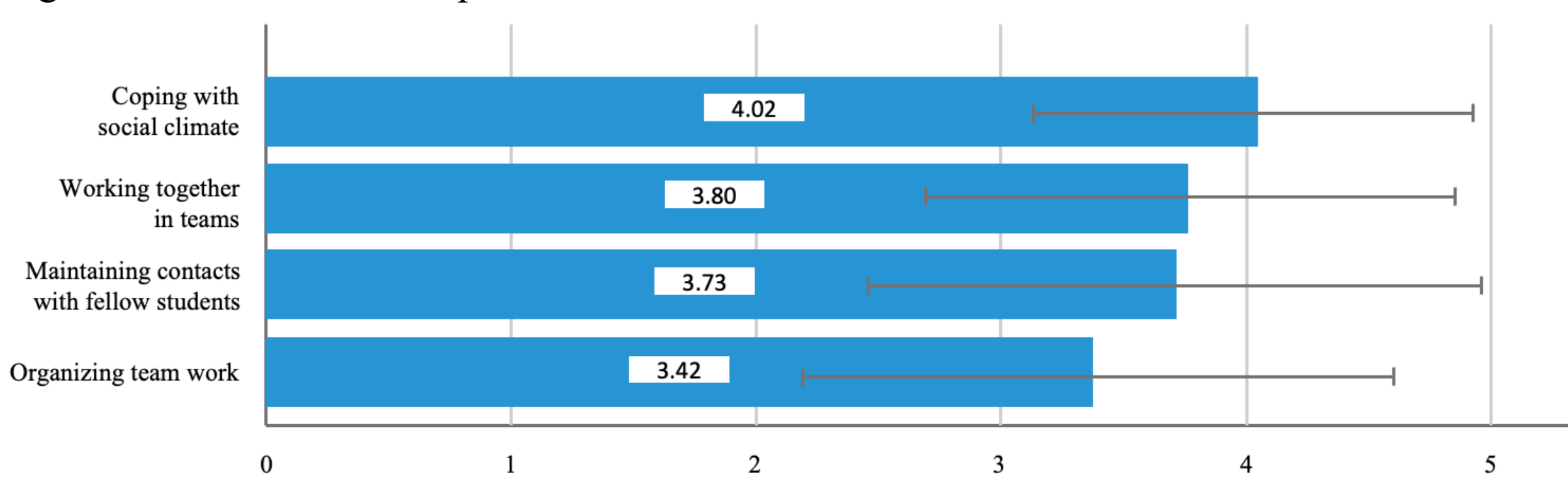

Responses were provided on a five-point Likert scale (1: very difficult, 2: difficult, 3: partly, 4: easy, 5: very easy; or doesn't fit for me/my studies). The questionnaire, including the original German items used, is included in the appendix. For each figure and table related to the questions in the AIS questionnaire, a brief translation of the German items into English is provided for enhanced clarity.

With a $SD$ above 1, maintaining contact with fellow students ($M = 3.73$, $SD = 1.18$) and organizing teamwork ($M = 3.42$, $SD = 1.14$) were rated with a wider range than the other two items.

Table 27: Contact and cooperation

| **Contact and cooperation** | *n* | *min* | *max* | *mean* | *sd* |
|---|---|---|---|---|---|
| Coping with social climate | 234 | 1 | 5 | 4.02 | .86 |
| Working together in teams | 227 | 1 | 5 | 3.80 | .98 |
| Maintaining contacts with fellow students | 239 | 1 | 5 | 3.73 | 1.18 |
| Organizing team work | 222 | 1 | 5 | 3.42 | 1.14 |
| Valid N (listwise) | 207 | | | | |

In terms of study organization, the participants perceived the creation of a personalized schedule to be the least difficult ($M = 3.92$, $SD = 1.03$), while the navigation of course offerings was rated as the most difficult study demand ($M$ = 3.24, $SD$ = 1.01), as can be seen in Tab. 28 and Fig. 12.

Figure 12: Study organization

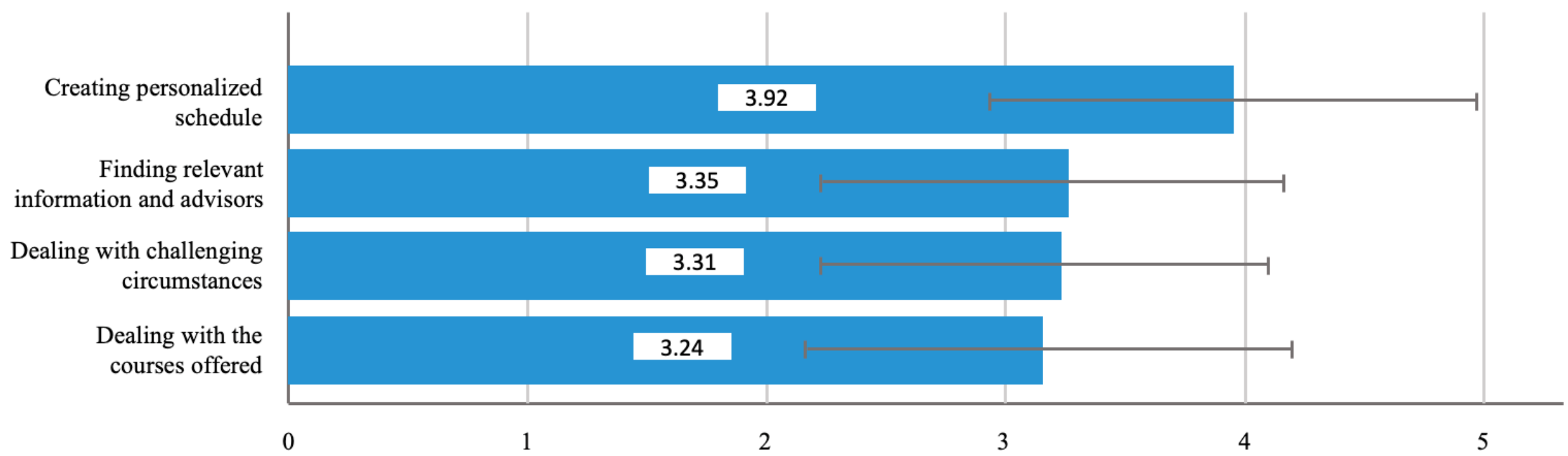

Responses were provided on a five-point Likert scale (1: very difficult, 2: difficult, 3: partly, 4: easy, 5: very easy; or does not fit for me/my studies). The questionnaire, including the original German items, is included in the appendix.

The $SD$ across all items in this subscale moves around 1. The two items with the highest – creating personalized schedule ($M$ = 3.92, $SD$ = 1.025) – and the lowest $SD$ – dealing with the courses offered ($M$ = 3.24, $SD$ = 1.013) – were rated with a wider range than the other two items.

Table 28: Study organization

| **Study organization** | *n* | *min* | *max* | *mean* | *sd* |
|---|---|---|---|---|---|
| Creating personalized schedule | 234 | 1 | 5 | 3.92 | 1.03 |
| Finding relevant information and advisors | 213 | 1 | 5 | 3.35 | .97 |
| Dealing with challenging circumstances | 227 | 1 | 5 | 3.31 | .93 |
| Dealing with courses offered | 241 | 1 | 5 | 3.24 | 1.01 |
| Valid N (listwise) | 197 | | | | |

Items assessing the pressure to perform in studies were also moderately rated, with means centered around the middle response option (see Tab. 29 and Fig. 13).

Figure 13: Performance pressure

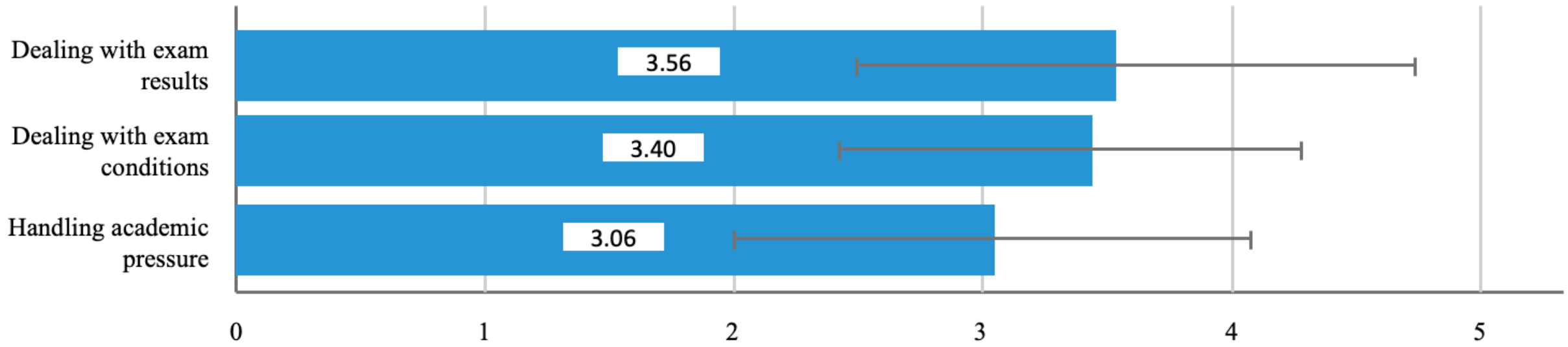

Answers were provided on a five-point Likert scale (1: very difficult, 2: difficult, 3: partly, 4: easy, 5: very easy; or does not fit for me/my studies). The entire scale, including the original German items used in the AIS-Questionnaire, is located in the appendix.

The average rating for dealing with exam results (*M* = 3.56, *SD* = 1.07) was between "partly" and "easy", while the average rating for handling academic pressure was between "partly" and "moderate" (*M* = 3.06, *SD* = 1.06).

Table 29: Performance pressure

| **Performance pressure** | *n* | *min* | *max* | *mean* | *sd* |
|---|---|---|---|---|---|
| Dealing with exam results | 235 | 1 | 5 | 3.56 | 1.07 |
| Dealing with exam Conditions | 237 | 1 | 5 | 3.40 | .97 |
| Handling academic pressure | 241 | 1 | 5 | 3.06 | 1.06 |
| Valid N (listwise) | 232 | | | | |

In terms of the learning activity requirements, there is a noticeable tendency towards the center of the response options. The highest mean value (*M* = 3.35, *SD* = 1.02) was observed for assessing personal resilience, whereas structuring learning activities were assessed with less difficulty on average (*M* = 3.02, *SD* = 1.06).

Figure 14: Learning activity

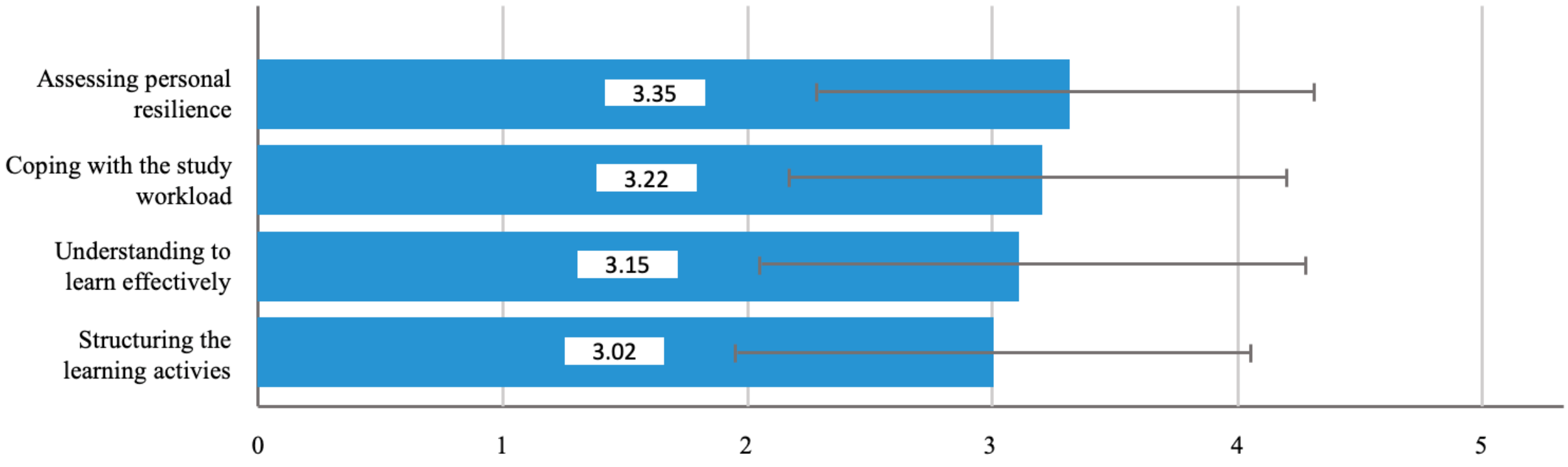

Responses were provided on a five-point Likert scale (1: very difficult, 2: difficult, 3: partly, 4: easy, 5: very easy; or does not fit for me/my studies). The entire scale, including the original German items used in the AIS-Questionnaire, is located in the appendix.

Similarly, the assessment of items in this subscale also ranges with values for the *SD* around 1, the highest value was calculated for the ability to understand and learn effectively ($M = 3.15$, $SD = 1.10$).

Table 30: Learning activity

| **Learning activity** | *n* | *min* | *max* | *mean* | *sd* |
|---|---|---|---|---|---|
| Assessing personal resilience | 239 | 1 | 5 | 3.35 | 1.02 |
| Coping with the study workload | 239 | 1 | 5 | 3.22 | .98 |
| Understanding to learn effectively | 241 | 1 | 5 | 3.15 | 1.10 |
| Structuring the learning activities | 240 | 1 | 5 | 3.02 | 1.06 |
| Valid N (listwise) | 238 | | | | |

Adapting academic approaches is perceived as the easiest study demand on the subscale for scientific mode ($M = 3.60$, $SD = 1.00$) whereas acquiring academic language skills ($M = 3.33$, $SD = 1.08$) and mastering academic methodologies ($M = 3.31$, $SD = 0.99$) received slightly lower average values (see Tab. 31 and Fig. 15).

Figure 15: Scientific mode

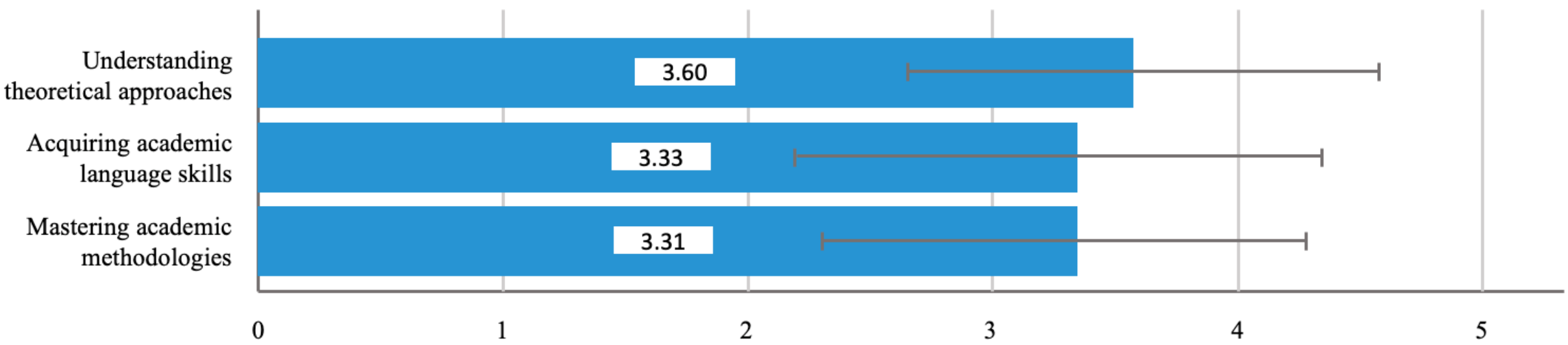

Responses were provided on a five-point Likert scale (1: very difficult, 2: difficult, 3: partly, 4: easy, 5: very easy; or does not fit for me/my studies). The complete scale, including the original German items used in the AIS-Questionnaire, is located in the appendix.

Table 31: Scientific mode

| **Scientific mode** | *n* | *min* | *max* | *mean* | *sd* |
|---|---|---|---|---|---|
| Understanding theoretical approaches | 235 | 1 | 5 | 3.60 | 1.00 |
| Acquiring academic language skills | 240 | 1 | 5 | 3.33 | 1.08 |
| Mastering academic methodologies | 236 | 1 | 5 | 3.31 | .99 |
| Valid N (listwise) | 231 | | | | |

The students' perceptions of study expectations were assessed as least difficult for the identification of personal interests and evaluation of programs (*M* = 3.78, *SD* = 0.99), somewhat more challenging for the establishment of links between theory and practice (*M* = 3.43, *SD* = 1.01), and most challenging for aligning course content with career aspirations (*M* = 3.01, *SD* = 1.10).

Figure 16: Study expectations

Responses were provided on a five-point Likert scale (1: very difficult, 2: difficult, 3: partly, 4: easy, 5: very easy; or does not fit for me/my studies). The entire scale, including the original German items used in the AIS-Questionnaire, is located in the appendix.

The item indicating the highest value on average is "identify own interests and match with study program" (*M* = 3.78, *SD* = 0.99), followed by "connecting theory and practice" (*M* = 3.43, *SD* = 1.01).

Table 32: Study expectations

| **Study expectations** | *n* | *min* | *max* | *mean* | *sd* |
|---|---|---|---|---|---|
| Identify own interests and match with study program | 227 | 1 | 5 | 3.78 | .99 |
| Connecting theory and practice | 238 | 1 | 5 | 3.43 | 1.01 |
| Combining the course of study with own interests | 234 | 1 | 5 | 3.01 | 1.10 |
| Valid N (listwise) | 220 | | | | |

The students rated contact and cooperation as the easiest requirement (*M* = 3.75, *SD* 0.77), while learning activity was rated the lowest (*M* = 3.18, *SD* 0.79). Overall, the results indicate a tendency towards the middle.

Figure 17: Values for subscales for sport students' perceptions of study demands

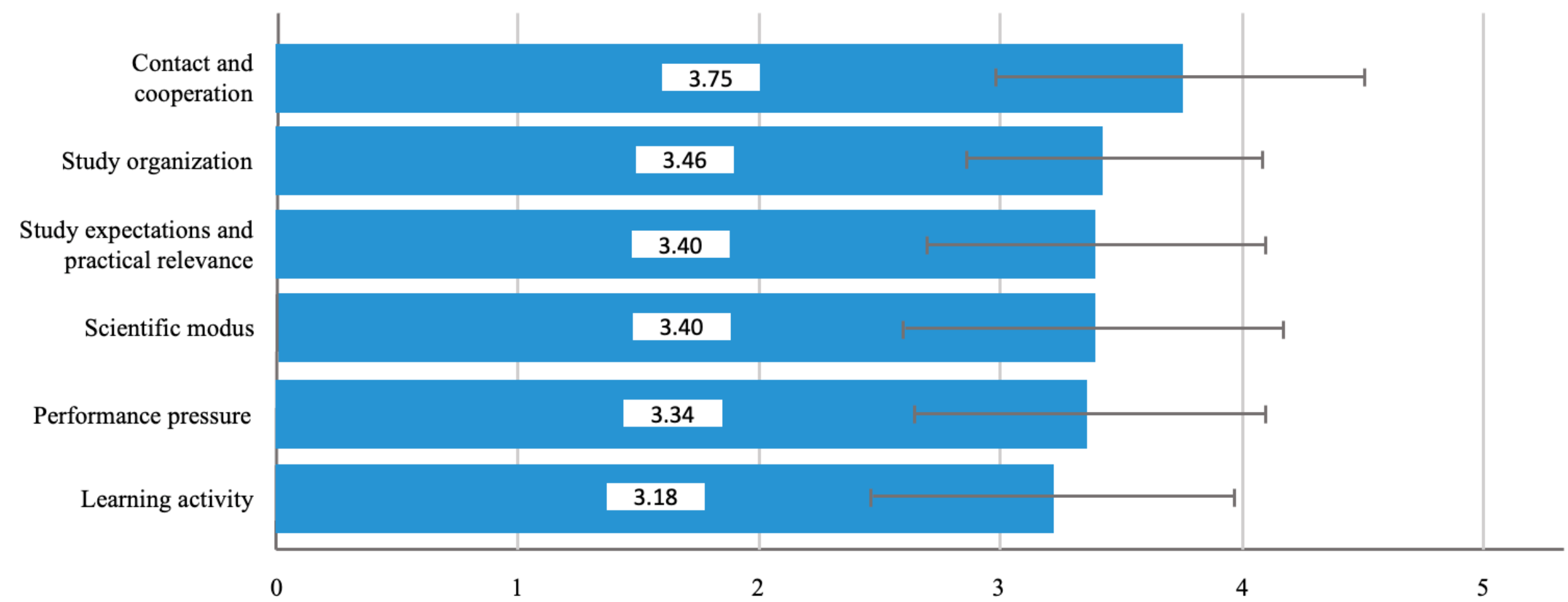

Responses were provided on a five-point Likert scale (1: very difficult, 2: difficult, 3: partly, 4: easy, 5: very easy; or does not fit for me/my studies). The complete scale, including the original German items utilized in the AIS Questionnaire, is presented in the appendix.

Table 33: Values for subscales for sport students' perceptions of study demands

| **Study demands** | *n* | *min* | *max* | *mean* | *sd* |
|---|---|---|---|---|---|
| Contact and cooperation | 242 | 1.50 | 5 | 3.75 | .77 |
| Study organization | 242 | 1.67 | 5 | 3.46 | .66 |
| Study expectations and practical relevance | 240 | 1 | 5 | 3.40 | .78 |
| Scientific modus | 241 | 1 | 5 | 3.40 | .87 |
| Performance pressure | 242 | 1 | 5 | 3.34 | .78 |
| Learning activity | 242 | 1 | 5 | 3.18 | .79 |
| Valid N (listwise) | 240 | | | | |

Students from the first two semesters indicated the highest level of difficulty on average ($M$ = 3.20, $SD$ = 0.57), whereas students in the 9th–10th semesters assessed the study demands with a tendency towards easier values ($M$ = 3.60, $SD$ = 0.52). With regard to the extreme values, it can be observed that none of the groups exhibited values below 2.10 or above 4.62. This indicates that, on average, students do not tend to rate their overall study demands as either very difficult (1) or very easy (5).

Table 34: Study progress and difficulty

| **Study Progress** | *n* | *min* | *max* | *mean* | *sd* |
|---|---|---|---|---|---|
| 1.-2. Semester | 23 | 2.10 | 4.48 | 3.20 | .57 |
| 3.-4. Semester | 47 | 2.14 | 4.48 | 3.38 | .59 |
| 5.-6. Semester | 38 | 2.24 | 4.43 | 3.46 | .45 |
| 7.-8. Semester | 40 | 2.38 | 4.33 | 3.40 | .47 |
| 9.-10. Semester | 31 | 2.81 | 4.57 | 3.60 | .52 |
| 11.-12. Semester | 23 | 2.84 | 4.60 | 3.49 | .40 |
| 13. Semester or higher | 14 | 2.47 | 4.30 | 3.47 | .48 |
| no answer | 2 | 2.52 | 2.95 | 2.74 | .30 |
| Missing values | 24 | 2.52 | 4.62 | 3.47 | .55 |

## 6.6 Overview of subtopics on student's perceptions of AI in studies

Figure 18 provides an overview of subthemes regarding students' perceptions of AI in sport studies, categorized based on the subthemes assessed in the AIS questionnaire. These included motivation, output quality, competency, future perspectives, and a general assessment of AI applications. To enhance students' overall understanding of AI in various subject areas, extremes, mean value, standard deviation, and variance were displayed. In order to facilitate interpretation, negatively formulated items were recoded in SPSS such that higher values indicate greater agreement with the positive issues represented by each subtheme listed below.

Figure 18: Student assessments of general AI topics

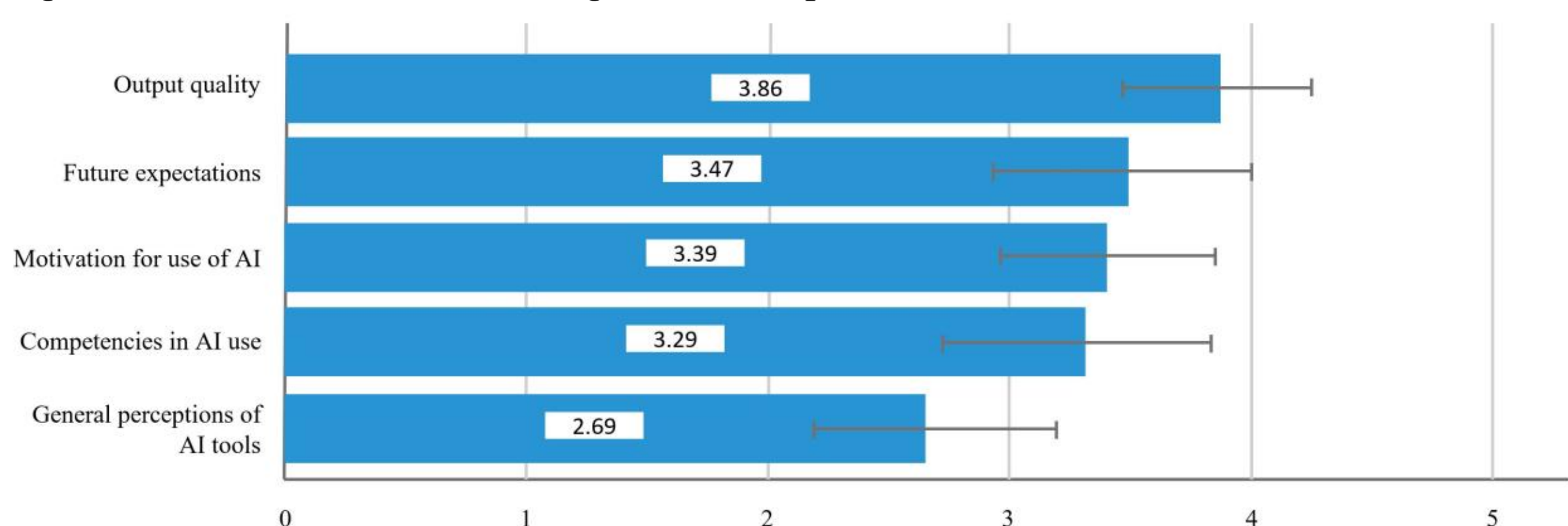

Responses were provided on a five-point Likert scale (1: strongly disagree, 2: disagree, 3: neither agree nor disagree, 4: agree, 5: strongly agree; or no response)

Table 35: Student assessments of general AI topics

| **Student ratings of general AI topics** | *n* | *min* | *max* | *mean* | *sd* |
|---|---|---|---|---|---|
| Output quality | 126 | 3.00 | 5.00 | 3.86 | .39 |
| Future expectations | 229 | 2.00 | 4.86 | 3.47 | .53 |
| Motivation for use of AI | 128 | 2.10 | 4.35 | 3.39 | .44 |
| Competencies in AI use | 122 | 2.00 | 4.75 | 3.30 | .57 |
| General perceptions of AI tools | 222 | 1.38 | 4.50 | 2.69 | .49 |
| Valid N (listwise) | 118 | | | | |

In general, students tended to provide responses that were situated towards the center of the scale. When these responses are compared to the study demands presented in Chapter 6.1, the lowest rating is received for general perceptions of AI tools ($M = 2.69$, $SD = 0.49$), indicating a rating between disagreement (2) and indecisiveness (3).

In response to both the AIS-specific question block and the study demands scale, students provided answers in the middle values. However, their assessment of study demands can be characterized as neither difficult nor easy. Their perceptions of AI, with its complex relevance for their studies, range around the indecisive option, ensuring no discrepancies between study demands and perceptions of AI arise.

Table 36: General requirements of the students

| **Study Demands** | *n* | *min* | *max* | *mean* | *sd* |
|---|---|---|---|---|---|
| Contact and cooperation | 242 | 1.50 | 5 | 3.75 | .77 |
| Study organization | 242 | 1.67 | 5 | 3.46 | .66 |
| Study expectations and practical relevance | 240 | 1 | 5 | 3.40 | .78 |
| Scientific modus | 241 | 1 | 5 | 3.40 | .87 |
| Performance pressure | 242 | 1 | 5 | 3.34 | .78 |
| Learning activity | 242 | 1 | 5 | 3.18 | .79 |
| Valid N (listwise) | 242 | | | | |

# 7
# Limitations

The goal of our study was to provide empirical insights into a university landscape undergoing fundamental change. As an ambitious private endeavor, we decided to make these findings freely available to researchers, students, lecturers, and other stakeholders interested in the current influence of AI in academia. The entire research process, including questionnaire development, outreach to sports departments, data collection, data analysis, and publication preparation, was conducted in our spare time. As a result, the present study is subject to certain limitations.

So far, the data collection has been limited to German students of sports science. Sport studies is a multidisciplinary field that integrates various elements from the social and natural sciences. In addition, a significant part of the curriculum is devoted to the education of future teachers. As a result, it is not currently possible to make direct comparisons with other countries, programs, or academic disciplines.

Moreover, the number of students participating in our study is relatively limited. While our study represents the largest sample of sports students in Germany surveyed about their use of AI in their studies, the participation of 262 students represents only about 1% of the total population of sports students in Germany.

This limitation is due less to student participation than to universities, which have received an increasing number of requests to participate in studies over the years. In response, some universities have implemented policies to manage external research requests. Many of the 62 sport departments we contacted responded to our questionnaire and thus contributed to our study.

Our study has so far focused exclusively on presenting quantitative results. Theoretical reflections or socio-cultural contextualizations of AI tools have not been included. Such explanatory and contextual research, exploring dimensions such as practices, discourses, and relationships in AI-centered human-computer interactions, is currently highly relevant. However, the development of such theoretical insights requires more time.

# 8
# Discussion

Our data supports common assumptions about AI, but also sheds light on new insights and paradoxical perspectives. First, it shows that students primarily associate "AI tools" with ChatGPT (98%), followed by Bing Chat (15.7%) and Google Bard (14.5%) (Tab. 10); and most students only use ChatGPT in their studies (62.6%). Among these, male students tend to use AI tools slightly more often than female students (Fig. 1).

In general, our findings indicate that sport students show a significant interest in integrating AI tools into their studies. They do not perceive AI as limiting their future engagement in sport activities and anticipate its increasing relevance in sport education, foreseeing a transformative impact on the traditional approach to studying sport (Fig. 10). They express confidence that the proliferation of AI tools in academia will not compromise critical thinking skills (Fig. 4) and support the integration of more AI-related topics into their curriculum, as well as the adoption of AI-based teaching methods by their lecturers (Fig. 5). They also have a positive outlook on the future societal impact of AI (Fig. 9).

Our data also shows that students primarily perceive AI tools as a "supplemental learning resource" to increase efficiency and intelligence in their academic endeavors, enabling them to save time, better understand complex concepts, and improve their overall academic performance (Fig. 4). At the same time, it is clear that students use AI tools primarily for creative tasks such as brainstorming, idea generation, and academic writing (Tab. 18). Secondarily, they are used as tools for activities such as literature research, concept understanding, or text translation. AI tools are rarely used for repetitive tasks such as composing emails, performing data calculations, or organizing workflows.

Most students agree that both lecturers and students should be proficient in the use of AI tools, while simultaneously rating the current skills of their instructors as very low (Fig. 5). Despite the high interest in acquiring relevant AI skills, only 11.1% of students report that their university offers courses on AI. 3.8% of students have already taken such a course, while the majority (64.9%) do not know if their university offers AI courses, followed by 19.1% of students who say that their university does not offer AI courses (see Tab. 13 & 15).

In terms of motivation, students recognize the benefits of AI tools; however, they face significant challenges in assessing the quality of knowledge generated by AI. While they see AI as a pioneering future technology, they believe that these tools can produce misinformation and distort perceptions of reality (Fig. 6). They believe that AI-generated knowledge may not always be accurate and needs additional verification. In this regard, potential plagiarism is seen as one of the major uncertainties associated with the use of AI tools (Fig. 4). This can be partly explained by the fact that most users are generally unaware of how AI tools generate knowledge and how they rely on the knowledge of others. Furthermore, the issue of plagiarism is well established in academic education, so most students are aware of the consequences of intellectual property theft. These findings also support the calls by researchers such as Sullivan et al. (2023) for increased research on integrity risks associated with the use of AI in higher education. Students tend to believe that AI has the potential to exacerbate threat scenarios, such as social crises and increasing social inequality (Fig. 9). At the same time, they are uncertain about AI's ability to help solve these crises (Fig. 8).

From these insights, a specific goal can be formulated: The current capabilities of AI tools have the potential to fundamentally transform existing research, study, and teaching methods. They enable new forms of accelerated, networked, and simultaneous knowledge generation that encompass creative and productive endeavors. Most students are aware of this potential. However, they lack the "internal compass" on how to use the results generated by AI while adhering to academic quality criteria such as reliability and protection of intellectual property.

In light of our empirical findings, we suggest that universities address this ambivalent situation by educating students on how to generate reliable information with AI and avoid misinformation. It is imperative that students have both practical and critical knowledge about the use and implications of a new technology that is rapidly spreading in academia. This will enable them to better understand the impact and also the limitations of a technology that only emerges through its practical use.

# References

Abdaljaleel, Maram, Muna Barakat, Mariam Alsanafi, Nesreen A Salim, Husam Abazid, Diana Malaeb, Ali Haider Mohammed, Bassam Abdul Rasool Hassan, Abdulrasool M Wayyes, and Sinan Subhi Farhan. 2024. "A multinational study on the factors influencing university students' attitudes and usage of ChatGPT". Scientific Reports 14: 1983. https://doi.org/10.1038/s41598-024-52549-8

Ansari, Aisha Naz, Sohail Ahmad, and Sadia Muzaffar Bhutta. 2023. "Mapping the global evidence around the use of ChatGPT in higher education: A systematic scoping review". Education and Information Technologies: 1-41. https://doi.org/10.1007/s10639-023-12223-4

Baidoo-Anu, David, and Leticia Owusu Ansah. 2023. "Education in the era of generative artificial intelligence (AI): Understanding the potential benefits of ChatGPT in promoting teaching and learning". Journal of AI 7, no. 1: 52-62. https://doi.org/10.61969/jai.1337500

Banh, Leonardo, and Gero Strobel. 2023. "Generative artificial intelligence". Electron Markets 33, no. 63. https://doi.org/10.1007/s12525-023-00680-1

Bolukbasi, Tolga, Kai-Wei Chang, James Y Zou, Venkatesh Saligrama, and Adam T Kalai. 2016. "Man is to computer programmer as woman is to homemaker? debiasing word embeddings". Advances in neural information processing systems 29. https://doi.org/10.48550/arXiv.1607.06520

Bonsu, Emmanuel Mensah, and Daniel Baffour-Koduah. 2023. "From the consumers' side: Determining students' perception and intention to use ChatGPT in Ghanaian higher education". Journal of Education, Society & Multiculturalism 4, no. 1: 1-29. https://doi.org/10.2478/jesm-2023-0001

Bozkurt, Aras, and Ramesh C. Sharma. 2023. "Generative AI and prompt engineering: The art of whispering to let the genie out of the algorithmic world". Asian Journal of Distance Education 18, no. 2: i-vii; https://doi.org/10.5281/zenodo.8174941

Buolamwini, Joy, and Timnit Gebru. 2018. "Gender Shades: Intersectional Accuracy Disparities in Commercial Gender Classification". Proceedings of Machine Learning Research, no. 81: 77-91. https://proceedings.mlr.press/v81/buolamwini18a/buolamwini18a.pdf

Burtell Matthew, and Helen Toner. 2024. "The Surprising Power of Next Word Prediction: Large Language Models Explained, Part 1". Center for Security and Emerging Technology, Georgetown University. https://cset.georgetown.edu/article/the-surprising-power-of-next-word-prediction-large-language-models-explained-part-1/

Burtell Matthew, and Helen Toner. 2023. "The Surprising Power of Next Word Prediction: Large Language Models Explained, Part 2". Center for Security and Emerging Technology, Georgetown University. https://cset.georgetown.edu/article/how-developers-steer-language-model-outputs-large-language-models-explained-part-2/

Cao, Yihan, Siyu Li, Yixin Liu, Zhiling Yan, Yutong Dai, Philip S. Yu, and Lichao Sun. 2023. "A Comprehensive Survey of AI-Generated Content (AIGC): A History of Generative AI from GAN to ChatGPT". arXiv preprint: 2303.04226. https://doi.org/10.48550/arXiv.2303.04226

Cascella, Marco, Jonathan Montomoli, Valentina Bellini, and Elena Bignami. 2023. "Evaluating the feasibility of ChatGPT in healthcare: An analysis of multiple clinical and research scenarios". Journal of Medical Systems 47, no. 1: 1–5; https://doi.org/10.1007/s10916-023-01925-4

Chan, Cecilia Ka Yuk. 2023. "A comprehensive AI policy education framework for university teaching and learning". International journal of educational technology in higher education 20, no. 38. https://doi.org/10.1186/s41239-023-00408-3

Chan, Cecilia Ka Yuk, and Wenjie Hu. 2023. "Students' Voices on Generative AI: Perceptions, Benefits, and Challenges in Higher Education". International Journal of Educational Technology in Higher Education 20, no. 43. https://doi.org/10.1186/s41239-023-00411-8

Chan, Cecilia Ka Yuk, and Wenxin Zhou. 2023. "An expectancy value theory (EVT) based instrument for measuring student perceptions of generative AI". Smart Learn. Environ. 10, no. 64. https://doi.org/10.1186/s40561-023-00284-4

Chmielewski, Christiana. 2018. "Self-Driving Cars and Rural Areas: The Potential for a Symbiotic Relationship." Journal of Law and Commerce 37, no. 1: 57–81. https://doi.org/10.5195/jlc.2018.153

Choi, Seongyune, Yeonju Jang, and Hyeoncheol Kim. 2023. "Influence of pedagogical beliefs and perceived trust on teachers' acceptance of educational artificial intelligence tools". International Journal of Human–Computer Interaction 39, no. 4: 910-22. https://doi.org/10.1080/10447318.2022.2049145

Dasborough, Marie T. 2023. "Awe-inspiring advancements in AI: The impact of ChatGPT on the field of organizational behavior." Journal of Organizational Behavior 44, no. 2: 177–179; https://doi.org/10.1002/job.2695

Dwivedi, Yogesh K, Anuj Sharma, Nripendra P Rana, Mihalis Giannakis, Pooja Goel, and Vincent Dutot. 2023. "Evolution of artificial intelligence research in Technological Forecasting and Social Change: Research topics, trends, and future directions". Technological Forecasting and Social Change 192: 122579. https://doi.org/10.1016/j.techfore.2023.122579

Frieder, Simon, Julius Berner, Philipp Petersen, and Thomas Lukasiewicz. 2023. "Large language models for mathematicians". arXiv preprint: 2312.04556. https://doi.org/10.48550/arXiv.2312.04556

Garrel, Jörg von, Jana Mayer, and Markus Mühlfeld. 2023. "Künstliche Intelligenz im Studium Eine quantitative Befragung von Studierenden zur Nutzung von ChatGPT & Co". Hochschule Darmstadt preprint. https://doi.org/10.48444/h_docs-pub-395

Geerling, Wayne, G Dirk Mateer, Jadrian Wooten, and Nikhil Damodaran. 2023. "ChatGPT has aced the test of understanding in college economics: Now what?". The American Economist 68, no. 2: 233-245. https://doi.org/10.1177/05694345231169654

Ghnemat, Rawan, Adnan Shaout, and Abrar M. Al-Sowi. 2022. "Higher Education Transformation for Artificial Intelligence Revolution: Transformation Framework". Int. J. Emerg. Technol. Learn. 17, no. 19: 224-41. https://doi.org/10.3991/ijet.v17i19.33309

Gilson, Aidan, Conrad W Safranek, Thomas Huang, Vimig Socrates, Ling Chi, Richard Andrew Taylor, and David Chartash. 2023. "How Does ChatGPT Perform on the United States Medical Licensing Examination (USMLE)? The Implications of Large Language Models for Medical Education and Knowledge Assessment". JMIR Medical Education 9: e45312. https://doi.org/10.2196/45312.

Guo, Lifang, Yin Yang, Hui Ding, Huiying Zheng, Hedan Yang, Junxiang Xie, Yong Li, Tong Lin, and Yiping Ge. 2022. "A deep learning-based hybrid artificial intelligence model for the detection and severity assessment of vitiligo lesions". Annals of Translational Medicine 10 (10): 590. https://doi.org/10.21037/atm-22-1738

Hsu, Tiffany, and Stuart A Thompson. 2023. Disinformation Researchers Raise Alarms About A.I. Chatbots. The New York Times. https:// www.nytimes.com/2023/02/08/technology/ai-chatbots-disinformation.html

Malik, Tegwen, Yogesh Dwivedi, Nir Kshetri, Laurie Hughes, Emma Louise Slade, Anand Jeyaraj, Arpan Kumar Kar, Abdullah M Baabdullah, Alex Koohang, and Vishnupriya Raghavan. 2023. "So what if ChatGPT wrote it?" Multidisciplinary perspectives on opportunities, challenges and implications of generative conversational AI for research, practice and policy. International Journal of Information Management, 71, 102642.

Intelligent. (2023, January 23). Nearly 1/3 college students have used ChatGPT on written assessments. Intelligent. https://www.intelligent.com/nearly-1-in-3-college-students-have-used-chatgpt-on-written-assignments/

Jänsch, Vanessa K., and Elke Bosse. 2018. Messinstrument für die Wahrnehmung von Studienanforderungen (MWS). Zusammenstellung sozialwissenschaftlicher Items und Skalen (ZIS). https://doi.org/10.6102/zis263

Kaplan, Andreas, and Michael Haenlein. 2019. "Siri, Siri, in my hand: Who's the fairest in the land? On the interpretations, illustrations, and implications of artificial intelligence". Business Horizons 62, no. 1: 15–25. https://doi.org/10.1016/j.bushor.2018.08.004

Karaca, Ozan, Ayhan S Caliskan, and Kadir Demir. 2021. "Medical artificial intelligence readiness scale for medical students (MAIRS-MS) - development, validity and reliability study". BMC MEDICAL EDUCATION 21: 112. https://doi.org/10.1186/s12909-021-02546-6

Kasneci, Enkelejda, Kathrin Seßler, Stefan Küchemann, Maria Bannert, Daryna Dementieva, Frank Fischer, Urs Gasser, Georg Groh, Stephan Günnemann, and Eyke Hüllermeier. 2023. "ChatGPT for good? On opportunities and challenges of large language models for education". Learning and Individual Differences 103: 102274. https://doi.org/10.1016/j.lindif.2023.102274

Khan, Ijaz, Abdul Rahim Ahmad, Nafaa Jabeur, and Mohammed Najah Mahdi. 2021. "An artificial intelligence approach to monitor student performance and devise preventive measures". Smart Learning Environments 8, no. 17: 1-18. https://doi.org/10.1186/s40561-021-00161-y

Kieslich, Kimon, Marco Lünich, Frank Marcinkowski, and Christopher Starke. 2019. "Hochschule der Zukunft: Einstellungen von Studierenden gegenüber Künstlicher Intelligenz an der Hochschule". Düsseldorf Institute for Internet und Democracy. https://doi.org/10.36200/0013

King, Michael R and ChatGPT. 2023. "A conversation on artificial intelligence, chatbots, and plagiarism in higher education". Cellular and molecular bioengineering 16: 1-2. https://doi.org/10.1007/s12195-022-00754-8

Kuhail, Mohammad Amin, Areej ElSayary, Shahbano Farooq, and Ahlam Alghamdi. 2022. "Exploring immersive learning experiences: A survey". Informatics 9, no. 4: 75. https://doi.org/10.3390/informatics9040075

Laupichler, Matthias Carl, Alexandra Aster, Jana Schirch, and Tobias Raupach. 2022. "Artificial intelligence literacy in higher and adult education: A scoping literature review". Computers and Education: Artificial Intelligence 3: 100101 https://doi.org/10.1016/j.caeai.2022.100101

Lim, Weng Marc, Asanka Gunasekara, Jessica Leigh Pallant, Jason Ian Pallant, and Ekaterina Pechenkina. 2023. "Generative AI and the future of education: Ragnarök or reformation? A paradoxical perspective from management educators". The International Journal of Management Education 21, no. 2: 100790. https://doi.org/10.1016/j.ijme.2023.100790

Lo, Chung Kwan. 2023. "What is the impact of ChatGPT on education? A rapid review of the literature". Education Sciences 13, no. 4: 410. https://doi.org/10.3390/educsci13040410

Mogali, Sreenivasulu Reddy. 2024. "Initial impressions of ChatGPT for anatomy education". Anatomical Sciences Education 17, no. 2: 444-47. https://doi.org/10.1002/ase.2261

Neumann, Michael, Maria Rauschenberger, and Eva-Maria Schön. 2023. "'We Need To Talk About ChatGPT': The Future of AI and Higher Education". IEEE/ACM 5th International Workshop on Software Engineering Education for the Next Generation (SEENG), Melbourne, Australia, 2023: 29-32. https://doi.org/10.1109/SEENG59157.2023.00010

Ngo, Thi Thuy An. 2023. "The perception by university students of the use of ChatGPT in education". International Journal of Emerging Technologies in Learning 18, no. 17: 4-19. https://doi.org/10.3991/ijet.v18i17.39019

Ooi, Keng-Boon, Garry Wei-Han Tan, Mostafa Al-Emran, Mohammed A. Al-Sharafi, Alexandru Capatina, Amrita Chakraborty, Yogesh K. Dwivedi, Tzu-Ling Huang, Arpan Kumar Kar, Voon-Hsien Lee, Xiu-Ming Loh, Adrian Micu, Patrick Mikalef, Emmanuel Mogaji, Neeraj Pandey, Ramakrishnan Raman, Nripendra P. Rana, Prianka Sarker, Anshuman Sharma, Ching-I Teng, Samuel Fosso Wamba, and Lai-Wan Wong. 2023. "The Potential of Generative Artificial Intelligence Across Disciplines: Perspectives and Future Directions". Journal of Computer Information Systems: 1-32. https://doi.org/10.1080/08874417.2023.2261010

Perkins, Mike. 2023. “Academic Integrity considerations of AI Large Language Models in the post-pandemic era: ChatGPT and beyond”. Journal of University Teaching & Learning Practice 20, no. 2: 07. https://doi.org/10.53761/1.20.02.07

Popenici, Stefan AD, and Sharon Kerr. 2017. “Exploring the impact of artificial intelligence on teaching and learning in higher education”. Research and Practice in Technology Enhanced Learning 12: 22. https://doi.org/10.1186/s41039-017-0062-8

Raman, Raghu, Santanu Mandal, Payel Das, Tavleen Kaur, JP Sanjanasri, and Prema Nedungadi. 2023. “University students as early adopters of ChatGPT: Innovation Diffusion Study”. Research Square preprint. https://doi.org/10.21203/rs.3.rs-2734142/v1.

Rammert, Werner. 2023. “Wie die Soziologie zur ‚Künstlichen Intelligenz‘ kam: Eine kurze Geschichte ihrer Beziehung“. In: Muhle, Florian (Ed.): “Soziale Robotik – Eine sozialwissenschaftliche Einführung”. De Gruyter, Berlin, Boston. Pp. 31–66. https://doi.org/10.1515/9783110714944-003

Rasul, Tareq, Sumesh Nair, Diane Kalendra, Mulyadi Robin, Fernando de Oliveira Santini., Ladeira, Wagner Junior Ladeira, Mingwei Sun, Ingrid Day, Raouf Ahmad Rather, and Liz Heathcote. 2023. “The role of ChatGPT in higher education: Benefits, challenges, and future research directions”. Journal of Applied Learning and Teaching 6, no. 1. https://doi.org/10.37074/jalt.2023.6.1.29

Sallam, Malik, Nesreen Salim, Muna Barakat, Kholoud Al-Mahzoum, Alaa B. Al-Tammemi, Diana Malaeb, Rabih Hallit, and Souheil Hallit. 2023. “Validation of a technology acceptance model-based scale TAME-ChatGPT on health students attitudes and usage of ChatGPT in Jordan”. JMIR Medical Education preprint. https://preprints.jmir.org/preprint/48254

Schmohl, Tobias, Alice Watanabe, and Kathrin Schelling. 2023. Künstliche Intelligenz in der Hochschulbildung. Bielefeld: transcript Verlag. https://doi.org/10.1515/9783839457696

Shoufan, Abdulhadi. 2023. “Exploring Students’ Perceptions of ChatGPT: Thematic Analysis and Follow-Up Survey”. IEEE Access 11: 38805–38818. https://doi.org/10.1109/ACCESS.2023.3268224

Sullivan, Miriam, Andrew Kelly, and Paul McLaughlan. 2023. “ChatGPT in higher education: Considerations for academic integrity and student learning”. Journal of Applied Learning & Teaching 6, no.1: 1-10. https://doi.org/10.37074/jalt.2023.6.1.17

Szabo, Attila. 2023. “ChatGPT is a breakthrough in science and education but fails a test in sports and exercise psychology”. Baltic Journal of Sport and Health Sciences 1, no. 128: 25-40. https://doi.org/10.33607/bjshs.v1i128.1341

Zawacki-Richter, Olaf, Victoria I Marín, Melissa Bond, and Franziska Gouverneur. 2019. “Systematic review of research on artificial intelligence applications in higher education– where are the educators?”. International Journal of Educational Technology in Higher Education 16: 39. https://doi.org/10.1186/s41239-019-0171-0

# Appendix

## AIS online questionnaire

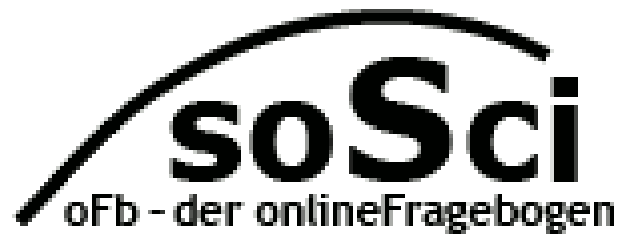

ki-im-sportstudium → base 31.08.2023, 09:51

**Seite 01**

Herzlich Willkommen zur Studie "Künstliche Intelligenz im Studium"!

Die Georg-August-Universität Göttingen untersucht zusammen mit dem Universitätsklinikum Hamburg-Eppendorf und der Universität Gent, welche Bedeutung Studierende KI-Anwendungen in ihrem Studium beimessen.

Wir wollen aus ihrer Sicht erfahren, ob sie bereits KI-Anwendungen verwenden, sie sich für den Einsatz in Lehre und Forschung grundsätzlich interessieren und welche Chancen und Risiken sie für die Zukunft in KI-Anwendungen sehen. In diesem Zuge fragen wir auch danach, wie sie mit den Anforderungen des Studiums umgehen und wie sie u.a. mit Prüfungsstress und Leistungsdruck zurechtkommen.

Unter KI-Anwendungen werden in der Studie Anwendungen verstanden, die textbasiert mittels Künstlicher Intelligenz verschiedene Inhalte wie z.B. Texte, Gliederungen, Bilder oder Konzepte generieren.

Das Ziel der Studie besteht darin, eine empirische Grundlage basierend auf Faktoren wie dem Nutzungsverhalten und der individuellen Bewertung von KI-Anwendungen zu schaffen, um aus dieser anschließend konkrete Handlungsempfehlungen für Studium, Lehre und Forschung herauszuarbeiten. Die Ergebnisse werden publiziert und sollen einen wichtigen Beitrag zur Sensibilisierung für das Thema und Verbesserung von Forschung und Lehre leisten. Zudem werden die Ergebnisse auch an die Studierenden kommuniziert.

Der Fragebogen umfasst KI bezogene Fragen sowie Fragen zum Studium und zur Lebenssituation. Die Beantwortung dauert etwa 10-15 Minuten.

Für den Erfolg der Studie ist es wichtig, dass Sie den Fragebogen vollständig ausfüllen.
Alle Daten werden anonym erhoben und streng vertraulich behandelt. Sie können keiner Person zugeordnet werden. Die Teilnahme an der Befragung ist vollständig freiwillig. Sie können Ihre Einwilligung jederzeit widerrufen, indem Sie die Beantwortung des Fragebogens abbrechen. Der Widerruf gilt nur für die Zukunft. Ihre Daten werden dann unverzüglich gelöscht.

Bei Fragen zur Studie steht Dr. Dennis Krämer als Projektleiter zur Verfügung (dennis.kraemer@uni-goettingen.de).

Vielen Dank für Ihre Teilnahme!
Dr. Dennis Krämer und Anja Bosold mit Team

---

**Seite 02**

**DA**

Die von Ihnen angegebenen Informationen werden in anonymisierter Form im Arbeitsbereich "Sport- und Gesundheitssoziologie" des Göttinger Instituts für Sportwissenschaften gespeichert. Sie sind nur den Projektmitgliedern zugänglich und werden ausschließlich zu Forschungszwecken verarbeitet. Es werden keine weiteren Informationen neben den von Ihnen angegebenen Antworten gespeichert. Es ist kein Zusammenhang zwischen Ihren personenbezogenen Daten und zu einem späteren Zeitpunkt veröffentlichten Forschungsergebnissen möglich.
Weitere Informationen zum Datenschutz entnehmen Sie bitte den Hinweisblättern zum Datenschutz gemäß DSGVO Art. 13 und Art. 14.

Mit Klick auf "Ich möchte freiwillig an dieser Studie teilnehmen" bestätigen Sie, dass Sie die Hinweise zum Datenschutz in Bezug auf die Verarbeitung Ihrer personenbezogenen Daten gelesen haben und an der Studie teilnehmen möchten.

**Seite 03**

**SA**

**1. Schätzen Sie ein, wie leicht bzw. schwer es Ihnen im vergangenen Studienjahr gefallen ist, mit den folgenden Anforderungen im Studium umzugehen.**

| | sehr schwer | eher schwer | teils-teils | eher leicht | sehr leicht | trifft auf mich/meinen Studiengang nicht zu |
|---|---|---|---|---|---|---|
| mit dem sozialen Klima im Studiengang zurechtzukommen (z.B. Konkurrenz aushalten) | ◯ | ◯ | ◯ | ◯ | ◯ | ◯ |
| mit dem vorhandenen Lehrangebot zurechtzukommen (z.B. ungünstige Termine, eingeschränkte Themenwahl) | ◯ | ◯ | ◯ | ◯ | ◯ | ◯ |
| mit Prüfungsergebnissen umzugehen (z.B. schlechte Noten) | ◯ | ◯ | ◯ | ◯ | ◯ | ◯ |
| zu erkennen, wie man sinnvoll lernt (z.B. die richtige Methode wählen) | ◯ | ◯ | ◯ | ◯ | ◯ | ◯ |
| sich die Wissenschaftssprache anzueignen (z.B. für Klausuren und Hausarbeiten) | ◯ | ◯ | ◯ | ◯ | ◯ | ◯ |
| einen eigenen Stundenplan zu erstellen (z.B. interessengeleitet Veranstaltungen wählen, Studienplan bei der Auswahl berücksichtigen) | ◯ | ◯ | ◯ | ◯ | ◯ | ◯ |
| mit Leistungsdruck umzugehen (z.B. mit Prüfungsstress oder eigenen Ansprüchen zurechtkommen) | ◯ | ◯ | ◯ | ◯ | ◯ | ◯ |
| Studieninhalte mit Berufsvorstellungen zu verbinden (z.B. Berufsperspektiven entwickeln, berufliche Relevanz der Inhalte erkennen) | ◯ | ◯ | ◯ | ◯ | ◯ | ◯ |
| Lernaktivitäten zeitlich sinnvoll zu strukturieren (z.B. Zeitpunkt und Dauer des Lernens) | ◯ | ◯ | ◯ | ◯ | ◯ | ◯ |
| Teamarbeit zu organisieren (z.B. Lerngruppen finden) | ◯ | ◯ | ◯ | ◯ | ◯ | ◯ |
| passende Informations- und Beratungsangebote zu finden (z.B. Ansprechpersonen finden) | ◯ | ◯ | ◯ | ◯ | ◯ | ◯ |
| eigene Interessen zu erkennen und die Studiengangswahl zu überprüfen (z.B. die Frage, ob der Studiengang zu einem passt) | ◯ | ◯ | ◯ | ◯ | ◯ | ◯ |
| die Menge an Lernstoff zu bewältigen (z.B. semesterbegleitende Aufgaben oder Lektüren) | ◯ | ◯ | ◯ | ◯ | ◯ | ◯ |
| sich auf die wissenschaftlichen Herangehensweisen einzustellen (z.B. unterschiedlicher Umgang mit Inhalten in Schule und Uni) | ◯ | ◯ | ◯ | ◯ | ◯ | ◯ |
| mit Prüfungsbedingungen zurechtzukommen (z.B. Benotungspraxis, Prüfungsdichte) | ◯ | ◯ | ◯ | ◯ | ◯ | ◯ |
| Bezüge zwischen Theorie und Praxis herzustellen (z.B. Anwendungsbeispiele finden) | ◯ | ◯ | ◯ | ◯ | ◯ | ◯ |
| Kontakte zu Mitstudierenden zu knüpfen (z.B. für Lerngruppen, Freizeit) | ◯ | ◯ | ◯ | ◯ | ◯ | ◯ |
| eigene Belastbarkeit einzuschätzen (z.B. Lernumfang, Erholungsbedarf) | ◯ | ◯ | ◯ | ◯ | ◯ | ◯ |
| mit ungünstigen Rahmenbedingungen umzugehen (z.B. Überfüllung, Zugangsbeschränkungen, Ausstattung) | ◯ | ◯ | ◯ | ◯ | ◯ | ◯ |
| wissenschaftliche Arbeitsweisen zu erlernen (z.B. wissenschaftliche Texte bearbeiten, eine Fragestellung entwickeln) | ◯ | ◯ | ◯ | ◯ | ◯ | ◯ |
| im Team zusammen zu arbeiten (z.B. gemeinsam Aufgaben bearbeiten, Referate vorbereiten) | ◯ | ◯ | ◯ | ◯ | ◯ | ◯ |

---

**Seite 04**
**N1**

**2. Haben Sie jemals KI-Anwendungen genutzt?**

Unter KI-Anwendungen werden in der Studie Anwendungen verstanden, die textbasiert mittels Künstlicher Intelligenz verschiedene Inhalte wie z.B. Texte, Gliederungen, Bilder oder Konzepte generieren.

◯ ja
◯ nein

**Seite 05**
**N2**

**3. Von welchen KI-Anwendungen haben Sie vor dieser Studie bereits gehört?**

- ☐ Albus
- ☐ Altered
- ☐ AutoRegex
- ☐ Bing Chat
- ☐ BRIA
- ☐ ChatGPT
- ☐ CopyAI
- ☐ FlikI
- ☐ GitHub Copilot
- ☐ Google Bard
- ☐ Jasper
- ☐ Krisp
- ☐ Lightning AI
- ☐ Midjourney
- ☐ Notion AI
- ☐ RocketAI
- ☐ Runway
- ☐ Slides AI
- ☐ SuperCreator
- ☐ Voicemod
- ☐ YouChat
- ☐ sonstige und zwar ____________
- ☐ keine

## Seite 06

**N3**

**4. Welche KI-Anwendungen haben Sie davon jemals im Sportstudium benutzt?**

- ☐ Albus
- ☐ Altered
- ☐ AutoRegex
- ☐ Bing Chat
- ☐ BRIA
- ☐ ChatGPT
- ☐ CopyAI
- ☐ Flikl
- ☐ GitHub Copilot
- ☐ Google Bard
- ☐ Jasper
- ☐ Krisp
- ☐ Lightning AI
- ☐ Midjourney
- ☐ Notion AI
- ☐ RocketAI
- ☐ Runway
- ☐ Slides AI
- ☐ SuperCreator
- ☐ Voicemod
- ☐ YouChat
- ☐ ⇉ sonstige und zwar ⇇
- ☐ keine

## Seite 07

**N4**

**5. An meiner Universität gibt es derzeit Kurse zu KI-Anwendungen.**

[Bitte auswählen] ⌄

**6. Ich habe bereits Kurse zu KI-Anwendungen an meiner Universität besucht.**

[Bitte auswählen] ⌄

**Seite 08**
**N5**

**7. Wie häufig nutzen Sie KI-Anwendungen im Sportstudium?**

[Bitte auswählen]

**8. Wie stark stimmen Sie der folgenden Aussage zu?**

| | stimme überhaupt nicht zu **1** | stimme nicht zu **2** | weder noch **3** | stimme zu **4** | stimme voll und ganz zu **5** |
|---|---|---|---|---|---|
| Ich habe Interesse, KI-Anwendungen im Sportstudium zu nutzen. | ○ | ○ | ○ | ○ | ○ |

**Seite 09**
**N6**

**9. Wofür haben Sie bereits KI-Anwendungen im Sportstudium genutzt?**

- ☐ Analyse (z.B. Rechnen mit Daten)
- ☐ E-Mails (z.B. Nachrichten formulieren)
- ☐ Ideenfindung (z.B. für ein Forschungsthema)
- ☐ Lernen (z.B. Verstehen von Konzepten)
- ☐ Planen (z.B. Erstellen eines Zeitplans)
- ☐ Recherchieren (z.B. Literatur finden)
- ☐ Schreiben (z.B. Hausarbeiten)
- ☐ Übersetzungen (z.B. von Englisch nach Deutsch)
- ☐ sonstiges und zwar ____________

### 10. Das Nutzen von KI-Anwendungen fördert meine Motivation in folgenden Bereichen

| | stimme überhaupt nicht zu<br>**1** | stimme nicht zu<br>**2** | weder noch<br>**3** | stimme zu<br>**4** | stimme voll und ganz zu<br>**5** |
|---|---|---|---|---|---|
| Analyse (z.B. Rechnen mit Daten) | ○ | ○ | ○ | ○ | ○ |
| E-Mails (z.B. Nachrichten formulieren) | ○ | ○ | ○ | ○ | ○ |
| Ideenfindung (z.B. für ein Forschungsthema) | ○ | ○ | ○ | ○ | ○ |
| Lernen (z.B. Verstehen von Konzepten) | ○ | ○ | ○ | ○ | ○ |
| Planen (z.B. Erstellen eines Zeitplans) | ○ | ○ | ○ | ○ | ○ |
| Recherchieren (z.B. Literatur finden) | ○ | ○ | ○ | ○ | ○ |
| Schreiben (z.B. Hausarbeiten) | ○ | ○ | ○ | ○ | ○ |
| Übersetzungen (z.B. von Englisch nach Deutsch) | ○ | ○ | ○ | ○ | ○ |
| ⇉ sonstiges und zwar ⇇ | ○ | ○ | ○ | ○ | ○ |

**Seite 11**

**MO**

**11. Bitte kreuzen Sie an, wie sehr Sie folgenden Aussagen zu Ihrer Motivation, KI-Anwendungen im Sportstudium zu nutzen, zustimmen.**

| | stimme überhaupt nicht zu | stimme nicht zu | weder noch | stimme zu | stimme voll und ganz zu | keine Angabe |
|---|---|---|---|---|---|---|
| Ich mache mir Sorgen, wer alles auf meine personenbezogenen Daten Zugriff hat, wenn ich KI-Anwendungen nutze. | ○ | ○ | ○ | ○ | ○ | ○ |
| KI-Anwendungen helfen mir dabei, schwierige Themen und Konzepte besser zu verstehen. | ○ | ○ | ○ | ○ | ○ | ○ |
| KI-Anwendungen helfen mir im Sportstudium bei der sportpraktischen Ausbildung. | ○ | ○ | ○ | ○ | ○ | ○ |
| KI-Anwendungen behindern die Problemlösungskompetenz. | ○ | ○ | ○ | ○ | ○ | ○ |
| Durch KI-Anwendungen kann ich meine Aufgaben im Sportstudium zeitsparender bewältigen. | ○ | ○ | ○ | ○ | ○ | ○ |
| KI-Anwendungen helfen mir im Sportstudium bei der theoretischen Ausbildung. | ○ | ○ | ○ | ○ | ○ | ○ |
| Ich kann KI-Anwendungen Fragen stellen, die ich den Dozierenden in Präsenz nicht stellen würde. | ○ | ○ | ○ | ○ | ○ | ○ |
| KI-Anwendungen helfen mir dabei, die Zusammenhänge von Sport und den theoretischen Teildisziplinen besser zu verstehen. | ○ | ○ | ○ | ○ | ○ | ○ |
| Durch den Einsatz von KI-Anwendungen im Sportstudium spare ich Geld. | ○ | ○ | ○ | ○ | ○ | ○ |
| KI-Anwendungen behindern die Fähigkeit im Team zu arbeiten. | ○ | ○ | ○ | ○ | ○ | ○ |
| KI-Anwendungen verbessern meine digitale Kompetenz. | ○ | ○ | ○ | ○ | ○ | ○ |
| KI-Anwendungen sind eine ergänzende Lernquelle. | ○ | ○ | ○ | ○ | ○ | ○ |
| Ich habe Bedenken, dass mir beim Nutzen von KI-Anwendungen Plagiat vorgeworfen wird. | ○ | ○ | ○ | ○ | ○ | ○ |
| KI-Anwendungen bieten mir einzigartige Perspektiven. | ○ | ○ | ○ | ○ | ○ | ○ |
| Der Einsatz von KI-Anwendungen im Sportstudium macht mir Spaß. | ○ | ○ | ○ | ○ | ○ | ○ |
| KI-Anwendungen unterstützen mich beim wissenschaftlichen Arbeiten. | ○ | ○ | ○ | ○ | ○ | ○ |
| KI-Anwendungen beeinträchtigen meine Fähigkeiten zum kritischen Denken. | ○ | ○ | ○ | ○ | ○ | ○ |
| KI-Anwendungen verbessern meine akademische Leistung. | ○ | ○ | ○ | ○ | ○ | ○ |
| KI-Anwendungen beeinträchtigen meine Kreativität. | ○ | ○ | ○ | ○ | ○ | ○ |
| Ich mache mir Sorgen, wer alles auf mein geistiges Eigentum Zugriff hat, wenn ich KI-Anwendungen nutze. | ○ | ○ | ○ | ○ | ○ | ○ |

**Seite 12**
**OG**

**12. Bitte kreuzen Sie an, wie sehr Sie folgenden Aussagen zur Güte der generierten Informationen von KI-Anwendungen zustimmen.**

| | Stimme überhaupt nicht zu | Stimme nicht zu | Weder noch | Stimme zu | Stimme voll und ganz zu | keine Angabe |
|---|---|---|---|---|---|---|
| Ich muss die von KI-Anwendungen produzierten Informationen prüfen, um sie für mein Studium nutzen zu können. | ○ | ○ | ○ | ○ | ○ | ○ |
| Die Unterscheidung zwischen selbst angefertigten Arbeiten und Arbeiten, die mittels KI-Anwendungen erstellt werden, ist schwierig. | ○ | ○ | ○ | ○ | ○ | ○ |
| KI-Anwendungen können Informationen generieren, die die Wirklichkeit verzerrt (d.h. nicht neutral) wiedergeben. | ○ | ○ | ○ | ○ | ○ | ○ |
| KI-Anwendungen können Fehlinformationen generieren. | ○ | ○ | ○ | ○ | ○ | ○ |
| KI-Anwendungen können voreingenommene Informationen produzieren. | ○ | ○ | ○ | ○ | ○ | ○ |
| Ich empfinde KI-Anwendungen als zuverlässig. | ○ | ○ | ○ | ○ | ○ | ○ |
| Mittels KI-Anwendungen erhalte ich die passenden Informationen, nach denen ich suche. | ○ | ○ | ○ | ○ | ○ | ○ |

**Seite 13**
**KO**

**13. Bitte kreuzen Sie an, wie sehr Sie folgenden Aussagen zur Rolle von Kompetenz im Umgang mit KI-Anwendungen im Sportstudium zustimmen.**

| | Stimme überhaupt nicht zu | Stimme nicht zu | Weder noch | Stimme zu | Stimme voll und ganz zu | keine Angabe |
|---|---|---|---|---|---|---|
| Dozierende sollten lernen, wie sie KI-Anwendungen in der Lehre einsetzen können. | ○ | ○ | ○ | ○ | ○ | ○ |
| Meine Dozierenden sind für den Umgang mit KI-Anwendungen ausreichend geschult. | ○ | ○ | ○ | ○ | ○ | ○ |
| Ich fühle mich für die Nutzung von KI-Anwendungen gerüstet. | ○ | ○ | ○ | ○ | ○ | ○ |
| Ich fühle mich ausreichend geschult, um mit KI-Anwendungen verschiedene Formen von Output (z.B. Text, Bild, ...) zu produzieren. | ○ | ○ | ○ | ○ | ○ | ○ |
| Studierende sollten lernen, wie sie KI-Anwendungen im Sportstudium nutzen können. | ○ | ○ | ○ | ○ | ○ | ○ |
| Ich habe eine Vorstellung davon, wie KI-Anwendungen Informationen generieren. | ○ | ○ | ○ | ○ | ○ | ○ |

**Seite 14**
**ZE**

**14. Bitte kreuzen Sie an, wie sehr Sie folgenden Aussagen zu Ihren Zukunftserwartungen in Bezug auf KI-Anwendungen im Sportstudium zustimmen.**

| | Stimme überhaupt nicht zu | Stimme nicht zu | Weder noch | Stimme zu | Stimme voll und ganz zu | keine Angabe |
|---|---|---|---|---|---|---|
| KI-Anwendungen gefährden die Arbeitsplätze von Menschen an Universitäten. | ○ | ○ | ○ | ○ | ○ | ○ |
| KI-Anwendungen werden sich negativ auf das Lernen auswirken. | ○ | ○ | ○ | ○ | ○ | ○ |
| Die Verbreitung von KI-Anwendungen wird dazu führen, dass man sich im Sportstudium weniger mit anderen austauscht. | ○ | ○ | ○ | ○ | ○ | ○ |
| KI-Anwendungen werden zukünftig eine immer größere Rolle im Sportstudium spielen. | ○ | ○ | ○ | ○ | ○ | ○ |
| KI-Anwendungen werden zukünftig das Sportstudium verändern. | ○ | ○ | ○ | ○ | ○ | ○ |
| Studierende werden Antworten und Lösungen für Aufgaben ohne große Anstrengung finden können. | ○ | ○ | ○ | ○ | ○ | ○ |
| Ich wünsche mir mehr Einsatz von KI-Anwendungen im Sportstudium. | ○ | ○ | ○ | ○ | ○ | ○ |

**15. Wo sehen Sie in der Zukunft die wichtigsten Einsatzgebiete von KI-Anwendungen im Studium?**

Beschreiben Sie mit eigenen Worten kurz, wo Sie zukünftig die wichtigsten Einsatzgebiete von KI-Anwendungen im Studium sehen.

Bitte geben Sie hier keine identifizierenden Daten ein.

**16. Bitte kreuzen Sie an, wie sehr Sie folgenden Aussagen zu Ihrer allgemeinen Einschätzung von KI-Anwendungen zustimmen.**

| | Stimme überhaupt nicht zu | Stimme nicht zu | Weder noch | Stimme zu | Stimme voll und ganz zu | keine Angabe |
|---|---|---|---|---|---|---|
| Ich mache mir Sorgen, dass KI-Anwendungen die soziale Ungleichheit verstärken werden. | ○ | ○ | ○ | ○ | ○ | ○ |
| Ich fühle mich unsicher, was den Einfluss von KI-Anwendungen auf mein zukünftiges Leben angeht. | ○ | ○ | ○ | ○ | ○ | ○ |
| KI-Anwendungen können die Welt verbessern. | ○ | ○ | ○ | ○ | ○ | ○ |
| Der Einfluss von KI-Anwendungen lässt sich nicht aufhalten. | ○ | ○ | ○ | ○ | ○ | ○ |
| Die Entwicklung von KI-Anwendungen lässt sich nicht aufhalten. | ○ | ○ | ○ | ○ | ○ | ○ |
| KI-Anwendungen sind für mich ein Schritt in Richtung Bevölkerungsüberwachung. | ○ | ○ | ○ | ○ | ○ | ○ |
| KI-Anwendungen können Gesellschaftskrisen verursachen. | ○ | ○ | ○ | ○ | ○ | ○ |
| KI-Anwendungen können Gesellschaftskrisen lösen. | ○ | ○ | ○ | ○ | ○ | ○ |

**17. Welchem Geschlecht ordnen Sie sich zu?**

[Bitte auswählen]

**18. Die Ergebnisse dieser Befragung werden auch für unterschiedliche Altersgruppen ausgewertet. Bitte nennen Sie dazu den Monat und das Jahr, in dem Sie geboren wurden.**

| | |
|---|---|
| Geburtsmonat | z.B. 01 für Januar |
| Geburtsjahr | z.B. 1999 |

**19. Welcher Nationalität gehören Sie gemäß Ihrer ersten Staatsbürgerschaft an?**

[Bitte auswählen]

**20. Haben Sie einen Migrationshintergrund?**

Eine Person hat einen Migrationshintergrund, wenn sie selbst oder mindestens ein Elternteil nicht mit deutscher Staatsangehörigkeit geboren wurde.

- ja
- nein
- weiß nicht
- keine Angabe

**21. Wie ist Ihr aktueller Familienstand?**

[Bitte auswählen]

**22. Haben Sie Kinder?**

Wenn ja, geben Sie bitte an, wie viele Kinder Sie haben.

- ◯ Ja
- ◯ Nein

---

- ◯ keine Angabe

**23. Wie ist Ihre derzeitige Wohnsituation?**

[Bitte auswählen]

**24. Wie oft treiben Sie Sport?**

| Keine sportliche Betätigung | Weniger als 1 Stunde in der Woche | Regelmäßig, 1-2 Stunden in der Woche | Regelmäßig, 2-4 Stunden in der Woche | Regelmäßig, mehr als 4 Stunden in der Woche | *keine Angabe* |
| --- | --- | --- | --- | --- | --- |

**25. Wie weit fortgeschritten sind Sie in Ihrem Studium?**

[Bitte auswählen]

**26. Welche beruflichen Ausbildungsabschlüsse haben Sie?**

Bitte nennen Sie alle Abschlüsse in dieser Liste, die Sie haben.

- ☐ Noch in beruflicher Ausbildung (Berufsvorbereitungsjahr, Auszubildende/r, Praktikant/-in, Student/-in)
- ☐ Schüler/-in und besuche eine berufsorientierte Aufbau-, Fachschule oder Ähnliches
- ☐ Keinen beruflichen Abschluss und bin nicht in beruflicher Ausbildung
- ☐ Beruflich-betriebliche Berufsausbildung (Lehre) abgeschlossen
- ☐ Berufsqualifizierenden Abschluss einer beruflich-schulischen Ausbildung (Berufsfachschule, Kollegschule)
- ☐ Vorbereitungsdienst für den mittleren Dienst in der öffentlichen Verwaltung
- ☐ Abschluss einer einjährigen Ausbildung an einer Schule des Gesundheitswesens
- ☐ Abschluss einer zwei- bis dreijährigen Ausbildung an einer Schule des Gesundheitswesens
- ☐ Abschluss einer Ausbildung zum Erzieher/zur Erzieherin
- ☐ Abschluss einer Fachschule der DDR
- ☐ Abschluss einer Fach-, Meister-, Technikerschule, Verwaltungs- und Wirtschaftsakademie oder Fachakademie
- ☐ Bachelor
- ☐ Diplom
- ☐ Master, Magister, Staatsexamen
- ☐ Promotion
- ☐ Einen anderen beruflichen Abschluss, und zwar: ____________

---

- ☐ keine Angabe

**Letzte Seite**

# Vielen Dank für Ihre Teilnahme!

Wir möchten uns ganz herzlich für Ihre Mithilfe bedanken.

Ihre Antworten wurden gespeichert, Sie können das Browser-Fenster nun schließen.

---

**Möchten Sie in Zukunft an interessanten und spannenden Online-Befragungen teilnehmen?**

Wir würden uns sehr freuen, wenn Sie Ihre E-Mail-Adresse für das SoSci Panel anmelden und damit wissenschaftliche Forschungsprojekte unterstützen.

E-Mail: [ ] Am Panel teilnehmen

Die Teilnahme am SoSci Panel ist freiwillig, unverbindlich und kann jederzeit widerrufen werden. Das SoSci Panel speichert Ihre E-Mail-Adresse nicht ohne Ihr Einverständnis, sendet Ihnen keine Werbung und gibt Ihre E-Mail-Adresse nicht an Dritte weiter.

Sie können das Browserfenster selbstverständlich auch schließen, ohne am SoSci Panel teilzunehmen.

---